\documentclass[1p]{elsarticle}
\usepackage{amssymb}
\usepackage{amsmath}
\usepackage{amsfonts}
\usepackage{graphics}
\usepackage{pstricks} %(pstcol) % pstricks commands
\usepackage{pst-node,pst-coil}
\usepackage{pst-optic}
\usepackage{pst-text}
\usepackage{hyperref}

\usepackage{tikz}

\newtheorem{theorem}{Theorem}

\newtheorem{proposition}[theorem]{Proposition}
\newtheorem{lemma}[theorem]{Lemma}
\newtheorem{observation}[theorem]{Observation}

\newtheorem{claim}{Claim}

\newproof{pf}{Proof}

\newcommand\ERASE[1]{}

\def\deg{{\rm deg}}

\def\chiE{\chi^e_{\Sigma}}
\def\chiT{\chi^t_{\Sigma}}
\def\EchiE{\overline{\chi^e_{\Sigma}}}
\def\EchiT{\overline{\chi^t_{\Sigma}}}

\begin{document}

\title{{\bf Equitable neighbour-sum-distinguishing\\ edge and total colourings}}

\author[labri,cnrs]{Olivier Baudon}
\author[agh]{Monika Pil\'sniak}
\author[agh]{Jakub Przyby{\l}o}
\author[labri,cnrs]{Mohammed Senhaji}
\author[labri,cnrs]{\'Eric Sopena}
\author[agh]{Mariusz Wo\'zniak\fnref{fn1}}

%\author[labri,cnrs]{Olivier Baudon\fnref{fn2}}
%\author[agh]{Monika Pil\'sniak\fnref{fn1}}
%\author[agh]{Jakub Przyby{\l}o\fnref{fn1}}
%\author[labri,cnrs]{\'Eric Sopena\fnref{fn2}}
%\author[labri,cnrs]{Mohammed Senhaji\fnref{fn2}}
%\author[agh]{Mariusz Wo\'zniak\fnref{fn1}}

%\fntext[fn1]{Partly supported by the Polish Ministry of Science and Higher Education.}
%\fntext[fn2]{ZZZZZZZZZZZZZZZZZZ.}
\address[agh]{AGH University of Science and Technology, al. A. Mickiewicza 30, 30-059 Krakow, Poland}
\address[labri]{Univ. Bordeaux, LaBRI, UMR5800, F-33400 Talence, France.}
\address[cnrs]{CNRS, LaBRI, UMR5800, F-33400 Talence, France.}
%
%\author{Monika Pil\'sniak\fnref{fn1}}
%{\small AGH University of Science and Technology}\\
%{\small Department of Discrete Mathematics}\\
%{\small al. Mickiewicza 30, 30-059  Krakow, Poland }\\
%{\small e-mail: {\tt \{pilsniak\}@agh.edu.pl}}\\
%
%\author{Jakub Przyby{\l}o\fnref{fn1}}
%\ead{przybylo@wms.mat.agh.edu.pl, phone: 048-12-617-46-38,  fax: 048-12-617-31-65}
%
%\fntext[fn2]{Partly supported by the Polish Ministry of Science and Higher Education.}
%
%
%\author{Mariusz Wo\'zniak, \thanks{The research partially supported by the Polish Ministry of Science and Higher Education}}
%{\small AGH University of Science and Technology}\\
%{\small Department of Discrete Mathematics}\\
%{\small al. Mickiewicza 30, 30-059  Krakow, Poland }\\
%{\small e-mail: {\tt \{mwozniak\}@agh.edu.pl}}\\
%
%

\begin{abstract}
With any (not necessarily proper) edge $k$-colouring $\gamma:E(G)\longrightarrow\{1,\dots,k\}$ of a graph $G$,
one can associate a vertex colouring $\sigma_{\gamma}$ given by $\sigma_{\gamma}(v)=\sum_{e\ni v}\gamma(e)$.
A neighbour-sum-distinguishing edge $k$-colouring is an edge colouring whose associated vertex colouring is proper.
The neighbour-sum-distinguishing index of a graph $G$ is then the smallest $k$ for which $G$ admits
a neighbour-sum-distinguishing edge $k$-colouring.
These notions naturally extends to total colourings of graphs that assign colours to both vertices and edges.

We study in this paper equitable neighbour-sum-distinguishing edge colourings and
%equitable neighbour-sum-distinguishing
total colourings, that is colourings $\gamma$ for which
the number of elements in any two colour classes of $\gamma$ differ by at most one.
We determine the equitable neighbour-sum-distinguishing index
of complete graphs, complete bipartite graphs and forests,
and the equitable neighbour-sum-distinguishing total chromatic number
of complete graphs and bipartite graphs.
\end{abstract}

\begin{keyword}
Equitable colouring
\sep Neighbour-sum-distinguishing edge colouring
\sep Neighbour-sum-distinguishing total colouring
%\sep 1--2--3 Conjecture
%\sep 1--2 Conjecture
\end{keyword}

\maketitle

\section{Introduction and statement of results}
\label{sec:introduction}

We consider undirected simple graphs and denote by $V(G)$ and $E(G)$ the sets of vertices and edges of a graph $G$, respectively.
We denote by $\deg_G(u)$, or simply $\deg(u)$ whenever the graph $G$ is clear from the context, the degree of a vertex $u$ in  $G$.
%XXXX TO BE COMPLETED

An \emph{edge $k$-colouring}, $k\ge 1$, of a graph $G$ is a mapping $\gamma:E(G)\longrightarrow\{1,\dots,k\}$.
Note that such edge colouring is not necessarily proper.
Each edge colouring  naturally induces a vertex colouring $\sigma_{\gamma}(v)$ given by
$$\sigma_{\gamma}(v)=\sum_{e\ni v}\gamma(e)$$
 for every $v\in V(G)$.
We will write $\sigma(v)$ instead of $\sigma_{\gamma}(v)$ whenever the edge colouring $\gamma$ is clear from the context.
We call the value $\gamma(e)$, $e\in E(G)$,  the \emph{colour of $e$} and the value $\sigma(v)$, $v\in V(G)$,  the \emph{sum at $v$}.
For an edge $uv\in E(G)$, we will say that $u$ and $v$ are \emph{in conflict} whenever $\sigma(u)=\sigma(v)$.
A \emph{neighbour-sum-distinguishing edge $k$-colouring} of $G$
(\emph{edge $k$-nsd-colouring} for short) is an edge $k$-colouring  $\gamma$ such that
for every edge $uv\in E(G)$, $\sigma_{\gamma}(u)\neq\sigma_{\gamma}(v)$.
Clearly, a graph admits an edge nsd-colouring if and only if it has no isolated edge.
The smallest $k$ for which $G$ admits an edge $k$-nsd-colouring is
the \emph{nsd-index} of $G$, denoted $\chiE(G)$.

A \emph{total $k$-colouring}, $k\ge 1$, of a graph $G$ is a mapping $\gamma_t:V(G)\cup E(G)\longrightarrow\{1,\dots,k\}$.
Again, note that we do not require a total colouring to be proper.
Similarly as above, each total colouring  naturally induces a vertex colouring $\sigma^T_{\gamma_t}(v)$ given by
$$\sigma^T_{\gamma_t}(v)=\gamma_t(v) + \sum_{e\ni v}\gamma_t(e)$$
 for every $v\in V(G)$.
We will write $\sigma^T(v)$ instead of $\sigma^T_{\gamma_t}(v)$ whenever the total colouring $\gamma_t$ is clear from the context.
We call the value $\gamma_t(e)$, $e\in E(G)$,  the \emph{colour of $e$}, the value $\gamma_t(v)$, $v\in V(G)$,  the \emph{colour of $v$}
 and the value $\sigma^T(u)$, $u\in V(G)$,  the \emph{sum at $u$}.
A \emph{total $k$-nsd-colouring} of $G$ is a total $k$-colouring  $\gamma_t$ such that
for every edge $uv\in E(G)$, $\sigma^T_{\gamma_t}(u)\neq\sigma^T_{\gamma_t}(v)$.
Clearly, every graph admits a total nsd-colouring.
The smallest $k$ for which $G$ admits a  total $k$-nsd-colouring is
the \emph{total nsd-chromatic number} of $G$, denoted $\chiT(G)$.

The study of edge nsd-colourings of graphs was initiated
by Karo\'nski, {\L}uczak and Thomason~\cite{123KLT}.
They  conjectured that every graph with no isolated edge admits an edge nsd-colouring  with three colours: $1,2,3$,
and proved it for $3$-colourable graphs.
Despite many efforts to tackle this conjecture, see e.g.~\cite{Louigi30,Louigi,123with13}, it is still an open question.
The best known result is due to Kalkowski, Karo\'nski and Pfender~\cite{KalKarPf123}, who proved that every graph
with no isolated edge admits an edge nsd-colouring  with five colours, $1,\ldots,5$.

Total nsd-colourings of graphs were introduced in~\cite{12Conjecture} by two of the authors, who %in~\cite{Prz08}.
%In~\cite{12Conjecture}, two of the authors
conjectured that every graph admits a
total $2$-nsd-colouring, and proved this fact for several graph families including in particular $3$-colourable, $4$-regular and complete graphs.
The best general upper bound is in this case due to Kalkowski, who proved in~\cite{Kalkowski12} sufficiency of integers $1,2,3$ for constructing
a  total nsd-colouring of any graph.

%We refer the reader interested by neighbour-distinguishing colourings to the survey by Seamone~\cite{Survey2012}.

\medskip

A colouring $\gamma$ with the elements of $\{1,\ldots,k\}$ is \emph{equitable} if the numbers of elements in any two colour classes differ by at most one,
that is $-1\le |\gamma^{-1}(i)| - |\gamma^{-1}(j)|\le 1$ for any two colours $i$ and $j$ with
$i,j\in\{1,\ldots,k\}$.
In 1964, Erd\"os~\cite{Erdos} conjectured that every graph with maximum degree ar most $r$ admits an equitable $(r+1)$-colouring. 
This conjecture was proved in 1970 by Hajnal and Szemer\'edi~\cite{HS70}
and is now known as Hajnal-Szemer\'edi Theorem.
A shorter proof of this theorem was given in 2008 by Kierstead and Kostochka~\cite{KK08}.
For a recent survey on equitable colourings, see~\cite{L13}.

In this paper, we study equitable edge and total nsd-colourings.
We will denote by $\EchiE(G)$
the \emph{equitable nsd-index} of $G$, that is
the smallest $k$ for which $G$ admits an equitable  edge $k$-nsd-colouring.
Similarly, we will denote by $\EchiT(G)$
the \emph{equitable  total nsd-chromatic number} of $G$, that is
the smallest $k$ for which $G$ admits an equitable  total $k$-nsd-colouring.

If $\gamma$ is an edge $k$-nsd-colouring of a graph $G$, then
the mapping $\gamma_t$ defined by $\gamma_t(v)=1$ for every $v\in V(G)$ and
$\gamma_t(e)=\gamma(e)$ for every $e\in E(G)$ is clearly a
 total $k$-nsd-colouring of $G$
since $\sigma_{\gamma_t}(v)=\sigma_{\gamma}(v)+1$ for every vertex $v\in V(G)$.
Hence, $\chiT(G)\le\chiE(G)$ for every graph $G$.
The same relation holds for equitable nsd-colourings:

\begin{proposition}
\label{prop:total_edges}
For every graph $G$ without isolated edges, $\EchiT(G) \leq \EchiE(G)$.
\end{proposition}

\begin{pf}
Let $\gamma$ be an equitable  edge $k$-nsd-colouring of $G$.
We will extend $\gamma$ to an equitable  total $k$-nsd-colouring $\gamma_t$ of $G$
with $\gamma_t(e)=\gamma(e)$ for every edge $e\in E(G)$.
We thus need to extend $\gamma$ to vertices in such a way
that no two adjacent vertices are in conflict and the colouring remains equitable.
We first order the vertices of $G$ as $v_1,v_2,\dots,v_n$, $n=|V(G)|$, in such a way that
\begin{equation}\label{eq:prop1}
\sigma_{\gamma}(v_1)\leq \sigma_{\gamma}(v_2)\leq \dots \leq \sigma_{\gamma}(v_n).
\end{equation}
Let $q_i=|\gamma^{-1}(i)|$ be the number of edges with colour $i$.
Moreover, let $q$ and $r$ be non-negative integers such that $r< k$ and
$|V(G)|+|E(G)|=qk+r$.
For $\gamma_t$ to be equitable, we then must have $r$ colour classes of order $q+1$
and $k-r$ colour classes of order $q$.
Let $C_q$ be any subset of $k-r$ colours from $\{1,\dots,k\}$ such that $q_i\le q$ for every $i\in C_q$,
and $C_{q+1}=\{1,\dots,k\}\setminus C_q$.
We will then colour $q'_i=q-q_i$ vertices with colour $i$ for each  $i\in C_q$ and $q'_j=q+1-q_j$ vertices with colour $j$
for each $j\in C_{q+1}$.
In order to produce a  total nsd-colouring, we will colour the vertices
according to the above defined order, and assign the colour 1 to the first $q'_1$ vertices, then colour 2
to the next $q'_2$ vertices and so one.
More formally, we let
\begin{equation}\label{eq:prop2}
\gamma_t(v_i)=\min\left\{j\ \Big|\ \sum_{p=1}^j q'_p\ge i\right\}.
\end{equation}
For every edge $v_iv_j$, $i<j$, we then have
$$\sigma_{\gamma_t}(v_i) = \sigma_{\gamma}(v_i) + \gamma_t(v_i) < \sigma_{\gamma}(v_j) + \gamma_t(v_j) = \sigma_{\gamma_t}(v_j)$$
since $\sigma_{\gamma}(v_i) < \sigma_{\gamma}(v_j)$ (by~(\ref{eq:prop1}), as $\gamma$ is neighbour-sum-distinguishing)
and $\gamma_t(v_i)\le \gamma_t(v_j)$ (by~(\ref{eq:prop2})).
The total colouring $\gamma_t$ is therefore an equitable total nsd-colouring,
and thus $\EchiT(G) \leq \EchiE(G)$.
\qed
\end{pf}

It is known that for every graph $G$ with no isolated edge, the  nsd-index
of $G$  is 3 if $G$ is a complete graph, see e.g.~\cite{CLWY11},
2~if $G$ is a  complete bipartite graph~\cite{LYZ11}
and at most~2 if $G$ is a forest~\cite{KNNSS12}.
Observe that we necessarily have $\chiT(G)\ge 2$, and thus $\chiE(G)\ge 2$, whenever
two adjacent vertices in $G$ have the same degree (in particular when $G$ is regular).

Concerning the equitable  nsd-index of these
graph classes, we will prove the following:

\begin{theorem}\label{TheoremEquitableSumCompleteGraph}
For every complete graph $K_n$ with $n\geq 3$, $n\neq 4$, $\EchiE(K_n)=3$, while $\EchiE(K_4)=4$.
\end{theorem}

\begin{theorem}\label{TheoremEquitableSumCompleteBipartiteGraph}
For every complete bipartite graph $K_{m,n}$ with $m=n=2$ or $m=n\geq 4$, $\EchiE(K_{m,n})=2$,
while $\EchiE(K_{3,3})=3$ and $\EchiE(K_{m,n})=1$ if $1\leq m<n$.
\end{theorem}

\begin{theorem}\label{TheoremEquitableSumForest}
For every forest $F$ with no isolated edge, $\EchiE(F)\leq 2$.
\end{theorem}

It was proved in~\cite{12Conjecture} that the total nsd-chromatic number of $G$
is~2 if $G$ is a complete graph of order $n\ge 2$,
and at most~2 if $G$ is bipartite.
We prove that the same
result holds for the
equitable  total nsd-chromatic number of bipartite graphs,
and that we need one more colour to produce an
equitable  total nsd-colouring of the complete graph $K_n$ whenever $n\ge 3$:

\begin{theorem}\label{TheoremEquitableSumTotalBipartite}
For every bipartite graph $G$, $\EchiT(G)\leq 2$.
\end{theorem}

\begin{theorem}\label{TheoremEquitableSumTotalComplete}
For every complete graph $K_n$ with $n\geq 3$, $\EchiT(K_n)=3$, while $\EchiT(K_2)=2$.
\end{theorem}

The proofs of Theorems~\ref{TheoremEquitableSumCompleteGraph}
to~\ref{TheoremEquitableSumTotalComplete} are given in the next sections.

\section{Proof of Theorem~\ref{TheoremEquitableSumCompleteGraph}}

We prove in this section that $\EchiE(K_n)=3$ for every $n\ge 3$, $n\neq 4$,
and that $\EchiE(K_4)=4$.
Recall that since $K_n$ is regular, we have $\EchiE(K_n)\ge 2$.
%Throughout the proof we will use only colours $1$, $2$ and $3$ to colour the edges.
We first introduce some definitions and preliminary results.

For any edge 3-colouring $\gamma$ of a graph $G$, we denote by $E_{\gamma}(i)$, $1\le i\le 3$, the set
of edges of $G$ with colour $i$. For every vertex $v$ of $G$, we denote by $d_{\gamma,i}(v)$ (or simply $d_i(v)$
when $\gamma$ is clear from the context), $1\le i\le 3$, the number of edges of $E_{\gamma}(i)$ that are incident with $v$.

Let $\sigma$ denote the vertex colouring of $G$ induced by $\gamma$,
%that is
%$$\sigma(v)=\sum_{e\ni v}\gamma(e),$$
%for every vertex $v\in V(G)$.
and $\overline{\sigma}$ denote the mean value of $\sigma$ on $V(G)$, that is
$$\overline{\sigma}=\frac{1}{|V(G)|}\sum_{v\in V(G)}\sigma(v).$$
The {\it $\sigma$-deviation} of a vertex $v$ is the value $\mu_{\sigma}(v)=\sigma(v)-\overline{\sigma}$.
For regular graphs, the deviation can be computed as follows:

\begin{lemma}\label{lemma:deviation-lambda}
Let $G$ be a $d$-regular graph, $\gamma$ be an edge 3-colouring of $G$
and $\sigma$ be the vertex colouring of $G$ induced by $\gamma$.
We then have $\sigma(v) = d_{\gamma,3}(v) - d_{\gamma,1}(v) + 2d$ for every vertex $v$ in $G$.
Moreover, if $|E_{\gamma}(1)|=|E_{\gamma}(3)|$, then 
$\overline{\sigma}=2d$
 and $\mu_{\sigma}(v)=d_{\gamma,3}(v)-d_{\gamma,1}(v)$ for every vertex $v$ in $G$.
\end{lemma}

\begin{pf}
We have
$$\begin{array}{rcl}
\sigma(v) & = & d_{\gamma,1}(v) + 2d_{\gamma,2}(v) + 3d_{\gamma,3}(v)\\
& = & d_{\gamma,3}(v) - d_{\gamma,1}(v) + 2(d_{\gamma,1}(v)+d_{\gamma,2}(v)+d_{\gamma,3}(v))\\
& = & d_{\gamma,3}(v) - d_{\gamma,1}(v) + 2d.
\end{array}.$$
Suppose now that $|E_{\gamma}(1)|=|E_{\gamma}(3)|$. We then have
$$
\begin{array}{rcl}
\sum_{v\in V(G)}\sigma(v) & = & \sum_{v\in V(G)}\left( d_{\gamma,3}(v) - d_{\gamma,1}(v) + 2d_G(v)\right)\\
& = & 2|E_{\gamma}(3)| - 2|E_{\gamma}(1)|  + 2d|V(G)|\\
& = & 2d|V(G)|.
\end{array}
$$
This gives $\overline{\sigma}=2d$.
Moreover, since
$\sigma(v) = d_{\gamma,3}(v) - d_{\gamma,1}(v) + 2d$,
we get $\mu_{\sigma}(v) = \sigma(v) - \overline{\sigma}=d_{\gamma,3}(v)-d_{\gamma,1}(v).$
\qed
\end{pf}

%Lemma~\ref{lemma:deviation-lambda} directly gives the following:
%
%\begin{corollary}\label{corollary:deviation}
%Let $G$ be a $d$-regular graph, $\gamma$ be an edge 3-colouring of $G$
%and $\sigma$ be the vertex colouring of $G$ induced by $\gamma$.
%If $\mu_{\sigma}$ defines a proper vertex colouring of $G$, then so does $\sigma$.
%\end{corollary}
%
% inutile et découle de la définition, pas du lemme...

%\begin{lemma}\label{lemma:deviation}
%Let $G$ be a $d$-regular graph, $\gamma$ be an edge 3-colouring of $G$ such that $|E_{\gamma}(1)|=|E_{\gamma}(3)|$,
%and $\sigma$ be the vertex colouring of $G$ induced by $\gamma$.
%
%We then have $\overline{\sigma}=2d$
% and $\mu_{\sigma}(v)=d_{\gamma,3}(v)-d_{\gamma,1}(v)$ for every vertex $v$ in $G$.
%\end{lemma}
%
%\begin{pf}
%Using Lemma~\ref{lemma:deviation-lambda}, we get
%$$
%\begin{array}{rcl}
%\sum_{v\in V(G)}\sigma(v) & = & \sum_{v\in V(G)}\left( d_{\gamma,3}(v) - d_{\gamma,1}(v) + 2d_G(v)\right)\\
%& = & 2|E_{\gamma}(3)| - 2|E_{\gamma}(1)|  + 2d|V(G)|\\
%& = & 2d|V(G)|.
%\end{array}
%$$
%This gives $\overline{\sigma}=2d$.
%Moreover, since
%$\sigma(v) = d_{\gamma,3}(v) - d_{\gamma,1}(v) + 2d$,
%we get $\mu_{\sigma}(v) = \sigma(v) - \overline{\sigma}=d_{\gamma,3}(v)-d_{\gamma,1}(v).$
%\qed
%\end{pf}

We say that an edge 3-colouring $\gamma$ of $G$ is {\it good} if the following conditions hold:
\begin{enumerate}
\item $|E_{\gamma}(1)| = |E_{\gamma}(3)|$,
%\item there exists a {\it 2-monochromatic triangle}, that is triangle $uvw$ in $G$ with $\gamma(uv)=\gamma(vw)=\gamma(wu)=2$,
\item for every vertex $v$ in $G$,
$$-\left\lfloor\frac{|V(G)|}{2}\right\rfloor \le \mu_{\sigma}(v) \le \left\lfloor\frac{|V(G)|}{2}\right\rfloor,$$
\item there exist two vertices $w_{min}^\gamma$ and $w_{max}^\gamma$ in $G$ such that
$$\mu_{\sigma}(w_{min}^\gamma) = -\left\lfloor\frac{|V(G)|}{2}\right\rfloor$$
and
$$\mu_{\sigma}(w_{max}^\gamma) = \left\lfloor\frac{|V(G)|}{2}\right\rfloor.$$
\end{enumerate}
where $\sigma$ denotes the vertex colouring of $G$ induced by $\gamma$.

\begin{lemma}\label{lemma:good-edge-colouring}
For every integer $n\ge 3$, if $\gamma$ is an equitable edge 3-colouring of the complete graph $K_n$
such that $|E_{\gamma}(1)| = |E_{\gamma}(3)|$,
then either
$$|E_{\gamma}(2)| = |E_{\gamma}(1)| = |E_{\gamma}(3)|$$
or
$$|E_{\gamma}(2)| = |E_{\gamma}(1)| + 1 = |E_{\gamma}(3)| + 1.$$
\end{lemma}

\begin{pf}
Since $|E(K_n)|=\frac{n(n-1)}{2}$,
if $n\equiv 0$ or $1 \pmod 3$, then $|E(K_n)|\equiv 0 \pmod 3$,
which implies $|E_{\gamma}(2)| = |E_{\gamma}(1)| = |E_{\gamma}(3)|$.
On the other hand, if $n\equiv 2 \pmod 3$, then $|E(K_n)| \equiv 1 \pmod 3$,
which implies $|E_{\gamma}(2)| = |E_{\gamma}(1)| + 1 = |E_{\gamma}(3)| + 1$.
\qed
\end{pf}

\medskip

We are now able to prove Theorem~\ref{TheoremEquitableSumCompleteGraph}.

\begin{pf}[of Theorem~\ref{TheoremEquitableSumCompleteGraph}]
We first prove a series of claims concerning equitable edge nsd-colourings
of $K_n$ for small values of $n$.

\begin{claim}\label{claim:complete-atleast3colours}
For every integer $n\ge 3$, $\overline{\chi_{\Sigma}^e}(K_n)\ge 3$.
\end{claim}

\begin{pf}
Assume to the contrary that $\gamma$ is an equitable  edge 2-nsd-colouring
of $K_n$ and let $\sigma$ be
the vertex colouring induced by $\gamma$.
We then have
$n-1\le\sigma(v)\le 2(n-1)$ for every vertex $v$ of $K_n$, hence,
since we need $n$ distinct values for $n$ vertices, there exist two vertices
$u$ and $v$ with $\sigma(u)=n-1$ and $\sigma(v)=2(n-1)$, in contradiction
with the colour of the edge $uv$.
\qed
\end{pf}

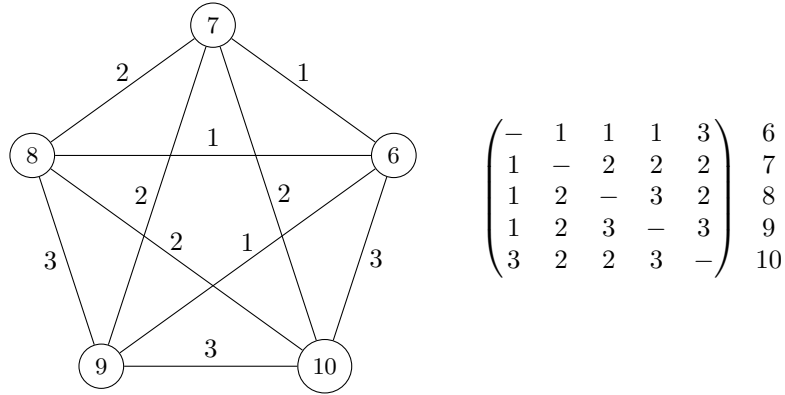
\begin{figure}[t]
\begin{center}
\begin{minipage}[c]{.45\linewidth}
\begin{tikzpicture}

\def \n {5}
\def \radius {2.5cm}
\def \nradius {0.6cm}
\def \margin {8} % margin in angles, depends on the radius

\foreach \s in {1,...,\n}
{
\pgfmathparse{5+\s} \let \z \pgfmathresult;
  \node (\s) at ({360/\n * (\s - 1) + 18}:\radius) [draw,circle, fill=none, minimum size=0.5cm] {\small \pgfmathprintnumber{\z}};

}

\draw (1)  -- (2)  node [midway, above, fill=none] {$1$};
\draw (1)  -- (3)  node [midway, above, fill=none] {$1$};
\draw (1)  -- (4)  node [midway, above, fill=none] {$1$};
\draw (1)  -- (5)  node [midway, right, fill=none] {$3$};
\draw (2)  -- (3)  node [midway, above, fill=none] {$2$};
\draw (2)  -- (4)  node [midway, left, fill=none] {$2$};
\draw (2)  -- (5)  node [midway, right, fill=none] {$2$};
\draw (3)  -- (4)  node [midway, left, fill=none] {$3$};
\draw (3)  -- (5)  node [midway, above, fill=none] {$2$};
\draw (4)  -- (5)  node [midway, above, fill=none] {$3$};
\end{tikzpicture}
\end{minipage}
\begin{minipage}[c]{.2\linewidth}
$
\begin{pmatrix}
-&1&1&1&3\\
1&-&2&2&2\\
1&2&-&3&2\\
1&2&3&-&3\\
3&2&2&3&-
\end{pmatrix}
\begin{array}{c}
6\\
7\\
8\\
9\\
10
\end{array}
$
\end{minipage}
\caption{\label{fig:K5Color} An equitable edge 3-nsd-colouring of $K_5$ and its
matrix representation.}
\end{center}
\end{figure}

\begin{figure}[t]
\begin{center}
$
\begin{pmatrix}
-&1&1&1&1&3\\
1&-&1&2&2&2\\
1&1&-&2&3&2\\
1&2&2&-&3&3\\
1&2&3&3&-&3\\
3&2&2&3&3&-
\end{pmatrix}
\begin{array}{r}
7\\
8\\
9\\
11\\
12\\
13
\end{array}
$
\caption{\label{fig:K6Color} Matrix representation of an equitable  edge 3-nsd-colouring of $K_6$.}
\end{center}
\end{figure}

\begin{claim}\label{claim:complete-K356}
For every integer $n\in\{3,5,6\}$, %$\overline{\chi_{\Sigma}^e}(K_n) = 3$.
there is a good equitable  edge 3-nsd-colouring of $K_n$.
\end{claim}

\begin{pf}
A good equitable  edge 3-nsd-colouring
of $K_3$ is obtained by colouring the edges of $K_3$ with colours 1, 2 and 3, respectively.

A good equitable edge 3-nsd-colouring
of $K_5$ is depicted in Figure~\ref{fig:K5Color},
together with its matrix representation. The value in row $i$ and column $j$ is the colour
of the edge $ij$. The sum at vertex $i$ is given at the end of row $i$.

The matrix representation of a good equitable  edge 3-nsd-colouring
of $K_6$ is given in Figure~\ref{fig:K6Color}.
\qed
\end{pf}

\begin{figure}[t]
\begin{center}
\begin{minipage}[c]{.45\linewidth}
\begin{tikzpicture}

\def \n {3}
\def \radius {2.5cm}
\def \nradius {0.6cm}
\def \margin {8} % margin in angles, depends on the radius

\foreach \s in {1,...,\n}
{
\pgfmathparse{5+\s} \let \z \pgfmathresult;
  \node (\s) at ({360/\n * (\s - 1) + 90}:\radius) [draw,circle, fill=none, minimum size=0.5cm] {\small \pgfmathprintnumber{\z}};

}

 \node (0) at (0,0) [draw,circle, fill=none, minimum size=0.5cm] {$5$};

\draw (1)  -- (2)  node [midway, left, fill=none] {$1$};
\draw (1)  -- (3)  node [midway, right, fill=none] {$3$};
\draw (0)  -- (1)  node [midway, left, fill=none] {$2$};
\draw (0)  -- (2)  node [midway, below, fill=none] {$2$};
\draw (2)  -- (3)  node [midway, below, fill=none] {$4$};
\draw (0)  -- (3)  node [midway, below, fill=none] {$1$};
\end{tikzpicture}
\end{minipage}
\begin{minipage}[c]{.2\linewidth}
$
\begin{pmatrix}
-&3&1&2\\
3&-&4&1\\
1&4&-&2\\
2&1&2&-
\end{pmatrix}
\begin{array}{c}
6\\
8\\
7\\
5
\end{array}
$
\end{minipage}

\caption{\label{fig:K4Color} An equitable  edge 4-nsd-colouring of $K_4$ and its matrix representation.}
\end{center}
\end{figure}

\begin{claim}\label{claim:complete-K4}
$\overline{\chi_{\Sigma}^e}(K_4) = 4$.
\end{claim}

\begin{pf}
Assume first that $\gamma$ is an equitable  edge 3-nsd-colouring
of $K_4$ and let $\sigma$ be the vertex colouring induced by $\gamma$.
Since $|E(K_4)|=6$, we necessarily have $|E_{\gamma}(1)|=|E_{\gamma}(2)|=|E_{\gamma}(3)|=2$
which implies $d_{\gamma,i}(v)\le 2$ for every $v\in V(K_4)$ and every $i$, $1\le i\le 3$.
Hence, $4\le\sigma(v)\le 8$ for every $v\in V(K_4)$.
On the other hand, we have
$\sum_{v\in V(K_4)}\sigma(v)=2\sum_{e\in E(K_4)}\gamma(e)=24.$

Let $V(K_4)=\{v_1,v_2,v_3,v_4\}$. To get a total sum of 24, we thus have, without loss of generality,
$\sigma(v_1) = 4$, $\sigma(v_2) = 5$, $\sigma(v_3) = 7$ and $\sigma(v_4) = 8$.
The edges incident with $v_1$ are thus coloured 1, 1 and 2, while the
edges incident with $v_4$ are coloured 3, 3 and 2. This implies $\gamma(v_1v_4)=2$
and thus $\gamma(v_1v_2)=1$, $\gamma(v_1v_3)=1$, $\gamma(v_2v_4)=3$ and $\gamma(v_3v_4)=3$,
hence $\gamma(v_2v_3)=2$. Thus, we finally get $\sigma(v_2)=\sigma(v_3)=6$, a contradiction.

Hence, $\overline{\chi_{\Sigma}^e}(K_4) > 3$. An equitable  edge 4-nsd-colouring
of $K_4$ is given in Figure~\ref{fig:K4Color},
together with its matrix representation.
\qed
\end{pf}

We will now prove that $\overline{\chi_{\Sigma}^e}(K_n) = 3$ for every $n\ge 7$, by induction on $n$.
More precisely, we will show that for every $n\ge 5$ any good
equitable  edge 3-nsd-colouring of $K_n$ can be extended to a good
equitable  edge 3-nsd-colouring of $K_{n+2}$.
Together with claims~\ref{claim:complete-atleast3colours} to~\ref{claim:complete-K4}, this will complete the proof.

Let $\gamma$ be a good equitable  edge 3-nsd-colouring of $K_n$, $n\ge 5$.
Suppose that $K_{n+2}$ is obtained from $K_n$ by adding two new vertices $u$ and $v$.
Let $S\subseteq V(K_n)$ be any fixed set of $\lfloor \frac{n}{2} \rfloor +1=\lfloor \frac{n+2}{2} \rfloor$ vertices
and $\overline{S}=V(K_n)\setminus S$ (we thus have $V(K_{n+2})=S\cup\overline{S}\cup\{u,v\}$).
%Let $u$ and $v$ be any two vertices of $K_{n+2}$.
We first define an edge 3-colouring $\gamma_0$ of $K_{n+2}$ as follows:
\begin{enumerate}
\item $\gamma_0(xy)=\gamma(xy)$ for every edge $xy$ with $x,y\in S\cup\overline{S}$,
\item $\gamma_0(uv)=2$,
\item for every vertex $x\in S$, $\gamma_0(ux)=1$ and $\gamma_0(vx)=3$,
\item for every vertex $y\in \overline{S}$, $\gamma_0(uy)=\gamma_0(vy)=2$.
\end{enumerate}

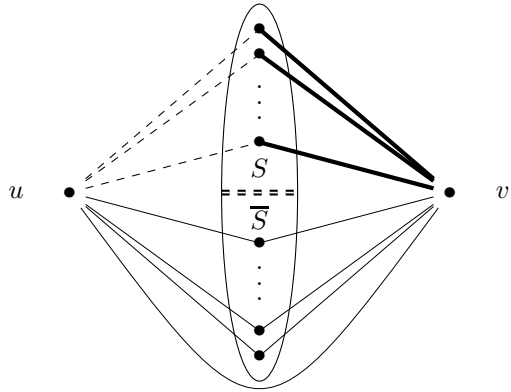
\begin{figure}[t]
\begin{center}
\begin{tikzpicture}

\def \n {3}
\def \radius {0.7cm}
\def \height {0.2cm}
\def \eL {2cm}
\def \el {0.5cm}
\def \margin {8} % margin in angles, depends on the radius

\draw [double, dashed, thick] (-\el,0) -- (\el,0);
\node (u) at ({-5*\el},0) [draw=none, fill=none] {$\bullet$};
\node[draw=none, fill=none] at ({-5*\el-\radius},0) {$u$};
\node (v) at ({5*\el},0) [draw=none, fill=none] {$\bullet$};
\node[draw=none, fill=none] at ({5*\el+\radius},0) {$v$};
\node[draw=none, fill=none] at (0,{\eL/6}) {$S$};
\node[draw=none, fill=none] at (0,{-\eL/6}) {$\overline{S}$};
\draw[rotate=90] (0, 0) ellipse (2.5cm and 0.5cm);
\foreach \s in {1,...,2}
{
	\node (h) at (0,{\eL-\s*(\eL/6)+0.5cm}) {$\bullet$};
	\node (l) at (0,{-\eL+\s*(\eL/6)-0.5cm}) {$\bullet$};
	\draw [dashed] (u) -- (h.center);
	\draw [ultra thick](v) -- (h.center);
	\draw [ultra thin] (u) -- (l.center);
	\draw [ultra thin] (v) -- (l.center);
}
\node (h) at (0,{\eL/3}) {$\bullet$};
\node (l) at (0,{-\eL/3}) {$\bullet$};
\draw [dashed] (u) -- (h.center);
\draw [ultra thick](v) -- (h.center);
\draw [ultra thin] (u) -- (l.center);
\draw [ultra thin] (v) -- (l.center);
\draw [ultra thin] (u) .. controls (0,{-1.7*\eL}) .. (v);
\node[draw=none, fill=none] at (0,{\height+\height+(\eL/2)}) {$.$};
\node[draw=none, fill=none] at (0,{\height+(\eL/2)}) {$.$};
\node[draw=none, fill=none] at (0,{(\eL/2)}) {$.$};
\node[draw=none, fill=none] at (0,{-(\eL/2)}) {$.$};
\node[draw=none, fill=none] at (0,{-\height-(\eL/2)}) {$.$};
\node[draw=none, fill=none] at (0,{-\height-\height-(\eL/2)}) {$.$};
\end{tikzpicture}
\caption{\label{fig:completecolor} The edge 3-colouring $\gamma_0$ of $K_{n+2}$.}
\end{center}
\end{figure}

The edge 3-colouring $\gamma_0$ is depicted on Figure~\ref{fig:completecolor}
(dashed, thin and thick edges represent edges with colour 1, 2 and 3, respectively).
This colouring is not equitable but is indeed good and neighbour-sum-distinguishing:

\begin{claim}\label{claim:gamma0-good}
The edge 3-colouring $\gamma_0$ is a good
 edge 3-nsd-colouring of $K_{n+2}$.
\end{claim}

\begin{pf}
Recall that $\gamma$ is a good equitable  edge 3-nsd-colouring of $K_n$.
We denote by $\sigma$ and $\sigma_0$ the vertex colourings induced by $\gamma$ and $\gamma_0$, respectively.

For each vertex $x\in S\cup\overline{S}$, the two edges $ux$ and $vx$ are either assigned colours 1 and 3 or both  assigned colour 2
by $\gamma_0$.
Therefore, for every vertex $x\in S\cup\overline{S}$, $\mu_{\sigma_0}(x) = \mu_{\sigma}(x)$, which implies
$-\lfloor \frac{n}{2} \rfloor\le \mu_{\sigma_0}(x) \le \lfloor \frac{n}{2} \rfloor$.
On the other hand,
$\mu_{\sigma_0}(u)=-(\lfloor \frac{n}{2} \rfloor +1)=-\lfloor \frac{n+2}{2}\rfloor$ and
$\mu_{\sigma_0}(v)=\lfloor \frac{n}{2} \rfloor + 1=\lfloor \frac{n+2}{2}\rfloor$.
Hence, all vertices of $K_{n+2}$ are assigned distinct values by $\mu_{\sigma_0}$.
Since $|E_{\gamma_0}(1)| = |E_{\gamma}(1)| + \lfloor \frac{n+2}{2} \rfloor$
and $|E_{\gamma_0}(3)| = |E_{\gamma}(3)| + \lfloor \frac{n+2}{2} \rfloor$,
so that $|E_{\gamma_0}(1)| = |E_{\gamma_0}(3)|$,
we thus get
by Lemma~\ref{lemma:deviation-lambda} %and~\ref{lemma:deviation} 
that $\gamma_0$ is a good  edge 3-nsd-colouring.
%
%%It remains to show that the edge colouring $\gamma_0$ is good.
%%
%%%The 2-monochromatic triangle of $K_n$ is still a 2-monochromatic triangle in $K_{n+2}$.
%%Finally, we already observed that $\mu_{\sigma_0}(u)=-\lfloor \frac{n+2}{2}\rfloor$ and
%%$\mu_{\sigma_0}(v)=\lfloor \frac{n+2}{2}\rfloor$,
%%and that $-\lfloor \frac{n}{2} \rfloor\le \mu_{\sigma_0}(x) \le \lfloor \frac{n}{2} \rfloor$
%%for every vertex $x\in S\cup\overline{S}$.
%%
%This completes the proof.
\qed
\end{pf}

While constructing $\gamma_0$ from $\gamma$, we added
$\lfloor \frac{n+2}{2} \rfloor$ edges with colour $1$, $\lfloor \frac{n+2}{2} \rfloor$ edges with colour $3$
and $2n+1 - 2\lfloor \frac{n+2}{2} \rfloor$ edges with colour $2$.
The edge colouring $\gamma_0$ is thus %(almost always) 
not necessarily equitable.
We will then modify the edge colouring $\gamma_0$
in order to obtain a good equitable edge colouring $\gamma_1$.
In order to do that, we need to recolour by 1 or 3 some edges which are coloured by 2,
say $p=2q$ such edges,
leading to an edge colouring $\gamma_1$ (with induced vertex colouring $\sigma_1$)
such that
$|E_{\gamma_1}(1)| = |E_{\gamma_0}(1)| + q$,
$|E_{\gamma_1}(3)| = |E_{\gamma_0}(3)| + q$ and
$|E_{\gamma_1}(2)| = |E_{\gamma_0}(2)| - 2q$
with either $|E_{\gamma_1}(2)| = |E_{\gamma_1}(1)|$ or $|E_{\gamma_1}(2)| = |E_{\gamma_1}(1)| + 1$
by Lemma~\ref{lemma:good-edge-colouring}.

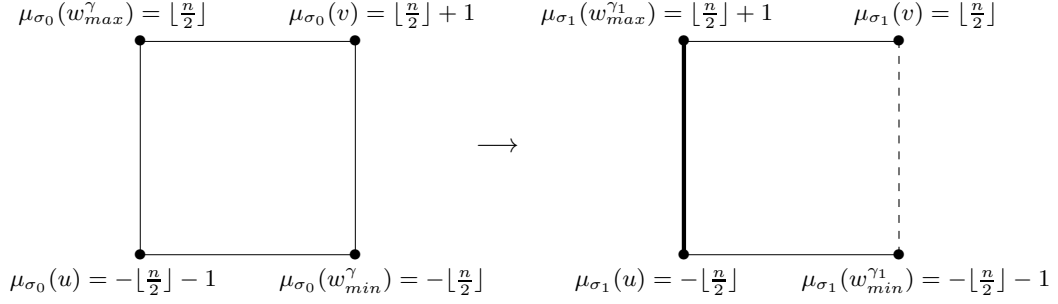
\begin{figure}[t]
\begin{center}
\begin{minipage}[c]{0.45\linewidth}
\begin{center}
\begin{tikzpicture}

\def \n {4}
\def \radius {2cm}
\def \margin {0.5cm}
\def \adjust {45}

\foreach \s in {1,...,\n}
{
	\node (\s) at ({\s*(360 / \n) + \adjust}:\radius) [draw=none, fill=none] {$\bullet$};
	\draw [ultra thin] ({\s*(360 / \n) + \adjust}:\radius) -- ({(\s+1)*(360 / \n) + \adjust}:\radius);
}
\node (v) at (45:{\radius+\margin}) [draw=none, fill=none] {\small $\mu_{\sigma_0}(v)= \lfloor \frac{n}{2} \rfloor+1$};
\node (w_{max}) at (135:{\radius+\margin}) [draw=none, fill=none] {\small $\mu_{\sigma_0}(w_{max}^{\gamma})= \lfloor \frac{n}{2} \rfloor$};
\node (u) at (225:{\radius+\margin}) [draw=none, fill=none] {\small $\mu_{\sigma_0}(u)= -\lfloor \frac{n}{2} \rfloor-1$};
\node (w_{max}) at (315:{\radius+\margin}) [draw=none, fill=none] {\small $\mu_{\sigma_0}(w_{min}^{\gamma})= -\lfloor \frac{n}{2} \rfloor$};

\end{tikzpicture}
\end{center}
\end{minipage}
\begin{minipage}[c]{0.05\linewidth}
\begin{center}
$\longrightarrow$
\end{center}
\end{minipage}
\begin{minipage}[c]{0.45\linewidth}
\begin{center}
\begin{tikzpicture}

\def \n {4}
\def \radius {2cm}
\def \margin {0.5cm}
\def \adjust {45}

\foreach \s in {1,...,\n}
{
	\node (\s) at ({\s*(360 / \n) +\adjust}:\radius) [draw=none, fill=none] {$\bullet$};
}
\node (v) at (45:{\radius+\margin}) [draw=none, fill=none] {\small $\mu_{\sigma_1}(v)= \lfloor \frac{n}{2} \rfloor$};
\node (w_{max}) at (135:{\radius+\margin}) [draw=none, fill=none] {\small $\mu_{\sigma_1}(w_{max}^{\gamma_1})= \lfloor \frac{n}{2} \rfloor +1$};
\node (u) at (225:{\radius+\margin}) [draw=none, fill=none] {\small $\mu_{\sigma_1}(u)= -\lfloor \frac{n}{2} \rfloor$};
\node (w_{min}) at (315:{\radius+\margin}) [draw=none, fill=none] {\small $\mu_{\sigma_1}(w_{min}^{\gamma_1})= -\lfloor \frac{n}{2} \rfloor-1$};

\draw[ultra thick] (135:\radius) -- (225:\radius);
\draw [ultra thin] (225:\radius) -- (315:\radius);
\draw[dashed] (315:\radius) -- (45:\radius);
\draw [ultra thin] (45:\radius) -- (135:\radius);

\end{tikzpicture}

\end{center}
\end{minipage}
\caption{\label{fig:recolouring-type-1} Recolouring of type $1$.}
\end{center}
\end{figure}

\begin{figure}[t]
\begin{center}
\begin{minipage}[c]{0.4\linewidth}
\begin{center}
\begin{tikzpicture}

\def \n {4}
\def \radius {2cm}
\def \margin {0.5cm}
\def \adjust {45}

\foreach \s in {1,...,\n}
{
	\node (\s) at ({\s*(360 / \n) +\adjust}:\radius) [draw=none, fill=none] {$\bullet$};
	\draw [ultra thin] ({\s*(360 / \n) +\adjust}:\radius) -- ({(\s+1)*(360 / \n) +\adjust}:\radius);
}
\def \s {1}
\node at ({\s*(360 / \n) +\adjust}:\radius+\margin) [draw=none, fill=none] {$y_i$};
\def \s {2}
\node at ({\s*(360 / \n) +\adjust}:\radius+\margin) [draw=none, fill=none] {$u$};
\def \s {3}
\node at ({\s*(360 / \n) +\adjust}:\radius+\margin) [draw=none, fill=none] {$x_i$};
\def \s {4}
\node at ({\s*(360 / \n) +\adjust}:\radius+\margin) [draw=none, fill=none] {$v$};
\end{tikzpicture}
\end{center}
\end{minipage}
\begin{minipage}[c]{0.1\linewidth}
\begin{center}
$\longrightarrow$
\end{center}
\end{minipage}
\begin{minipage}[c]{0.4\linewidth}
\begin{center}
\begin{tikzpicture}
\def \n {4}
\def \radius {2cm}
\def \margin {0.5cm}
\def \adjust {45}
\foreach \s in {1,...,\n}
{
	\node (\s) at ({\s*(360 / \n) + \adjust}:\radius) [draw=none, fill=none] {$\bullet$};
}
\def \s {1}
\draw [ultra thick]  ({\s*(360 / \n) + \adjust}:\radius) -- ({(\s+1)*(360 / \n) +\adjust}:\radius);
\node at ({\s*(360 / \n) +\adjust}:\radius+\margin) [draw=none, fill=none] {$y_i$};
\def \s {2}
\draw [dashed] ({\s*(360 / \n) + \adjust}:\radius) -- ({(\s+1)*(360 / \n) +\adjust}:\radius);
\node at ({\s*(360 / \n) +\adjust}:\radius+\margin) [draw=none, fill=none] {$u$};
\def \s {3}
\draw [ultra thick] ({\s*(360 / \n) + \adjust}:\radius) -- ({(\s+1)*(360 / \n) +\adjust}:\radius);
\node at ({\s*(360 / \n) +\adjust}:\radius+\margin) [draw=none, fill=none] {$x_i$};
\def \s {4}
\draw [dashed] ({\s*(360 / \n) + \adjust}:\radius) -- ({(\s+1)*(360 / \n) +\adjust}:\radius);
\node at ({\s*(360 / \n) +\adjust}:\radius+\margin) [draw=none, fill=none] {$v$};
\end{tikzpicture}
\end{center}
\end{minipage}
\caption{\label{fig:recolouring-type-2} Recolouring of type $2$.}
\end{center}
\end{figure}
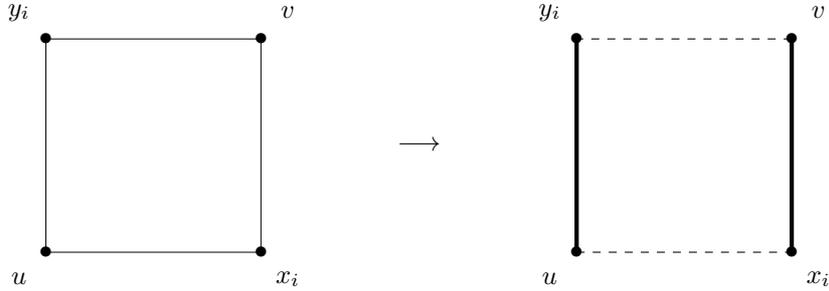

The edge colouring $\gamma_1$ will be produced using two types of recolourings,
both involving edges incident with $u$ or $v$, described as follows:
\begin{itemize}
\item {\it Recolouring of type 1} (see Figure~\ref{fig:recolouring-type-1}):
Let $w_{min}^{\gamma}$ and $w_{max}^{\gamma}$ denote the (unique) two vertices
such that $\mu_{\sigma_0}(w_{min}^{\gamma})=-\lfloor\frac{n}{2}\rfloor$ ($=\mu_{\sigma}(w_{min}^{\gamma})$) and
$\mu_{\sigma_0}(w_{max}^{\gamma})=\lfloor\frac{n}{2}\rfloor$ ($=\mu_{\sigma}(w_{max}^{\gamma})$).
If $uw_{min}^{\gamma}vw_{max}^{\gamma}$ is a 2-monochromatic 4-cycle,
then recolour with 1
the edge $vw_{min}^{\gamma}$ and recolour with 3 the edge $uw_{max}^{\gamma}$.
Note that the deviations of $u$ and $w_{min}^{\gamma}$,
and of $v$ and $w_{max}^{\gamma}$ have been switched, so that
$w_{min}^{\gamma_1}=w_{min}^{\gamma}$ and $w_{max}^{\gamma_1}=w_{max}^{\gamma}$.

\item {\it Recolouring of type 2} (see Figure~\ref{fig:recolouring-type-2}):
If the set of pairs of vertices $\{(x_i,y_i)\}_{1\le i\le k}$, $k\ge 1$,
is such that $ux_ivy_i$ is a 2-monochromatic 4-cycle for every $i$, $1\le i\le k$, then recolour with 1
all edges $ux_i$ and $vy_i$ and recolour with 3 all edges $uy_i$ and $vx_i$.
Note that the deviation of any of these $2k+2$ vertices remains unchanged.

\end{itemize}

Recall that we need to recolour $p=2q$ edges which are coloured by 2, $q$ of them with colour 1
and the $q$ others with colour 3.
If $q=1$, since $n\ge 5$, we can ensure that the chosen set $S$ contains none of the vertices
$w_{min}^{\gamma}$ and $w_{max}^{\gamma}$. By doing so and then applying the recolouring of type~1,
we obtain an edge colouring $\gamma_1$ such that:
\begin{enumerate}
\item $|E_{\gamma_1}(1)| = |E_{\gamma_1}(3)|$ and either $|E_{\gamma_1}(2)| = |E_{\gamma_1}(1)|$
or $|E_{\gamma_1}(2)| = |E_{\gamma_1}(1)| + 1$,
%\item since $n\ge 7$, there exists a vertex $w\in\overline{S}\setminus\{w_{min}^{\gamma_0},w_{max}^{\gamma_0}\}$ and
%$uvw$ is a 2-monochromatic triangle,
\item for every vertex $x\in V(K_{n+2})$, $-\lfloor\frac{n+2}{2}\rfloor \le \sigma_1(x) \le \lfloor\frac{n+2}{2}\rfloor$,
\item $w_{min}^{\gamma_1}=w_{min}^{\gamma}$ and $w_{max}^{\gamma_1}=w_{max}^{\gamma}$.
\end{enumerate}
Hence, $\gamma_1$ is
an equitable good  edge 3-nsd-colouring of $K_{n+2}$ and we are done.

Assume from now on that $q\ge 2$.
Since $\gamma$ is an equitable good  edge 3-nsd-colouring of $K_{n}$,
we know by Lemma~\ref{lemma:good-edge-colouring} that
$$|E_{\gamma}(1)| = |E_{\gamma}(3)| = r,\ \mbox{and}\ r \le |E_{\gamma}(2)| \le r + 1,$$
with $r = \lfloor\frac{n(n-1)}{6}\rfloor$.
As observed before, considering the way the edge colouring $\gamma_0$ has been
constructed (see Figure~\ref{fig:completecolor}), we also have
$$|E_{\gamma_0}(1)| = |E_{\gamma_0}(3)| = r + \left\lfloor\frac{n+2}{2}\right\rfloor,$$
and
$$r + 2n+1 - 2\left\lfloor\frac{n+2}{2}\right\rfloor \le |E_{\gamma_0}(2)| \le r + 2n+1 - 2\left\lfloor\frac{n+2}{2}\right\rfloor + 1.$$
Again by Lemma~\ref{lemma:good-edge-colouring}, in order to be an
equitable good  edge 3-nsd-colouring of $K_{n+2}$,
the edge colouring $\gamma_1$ must be such that
$$|E_{\gamma_1}(1)| = |E_{\gamma_1}(3)| = r + \left\lfloor\frac{n+2}{2}\right\rfloor + q,$$
and
\begin{equation}\label{eq:1}
r + \left\lfloor\frac{n+2}{2}\right\rfloor + q \le |E_{\gamma_1}(2)| \le r + \left\lfloor\frac{n+2}{2}\right\rfloor + q + 1.
\end{equation}
On the other hand, since $\gamma_1$ has been obtained by recolouring $2q$ edges which were coloured with colour 2 by $\gamma_0$,
we also have
\begin{equation}\label{eq:2}
r + 2n+1 - 2\left\lfloor\frac{n+2}{2}\right\rfloor - 2q \le |E_{\gamma_1}(2)| \le r + 2n+1 - 2\left\lfloor\frac{n+2}{2}\right\rfloor - 2q + 1.
\end{equation}
Combining~\eqref{eq:1} and~\eqref{eq:2}, we get
$$r + \left\lfloor\frac{n+2}{2}\right\rfloor + q \le r + 2n+1 - 2\left\lfloor\frac{n+2}{2}\right\rfloor - 2q + 1,$$
%$$r + 2n - 2\left\lfloor\frac{n+2}{2}\right\rfloor - 2q \le r + \left\lfloor\frac{n+2}{2}\right\rfloor + q + 1,$$
which gives
$$3q \le 2n+1 - 3\left\lfloor\frac{n+2}{2}\right\rfloor + 1.$$

Since $q\ge 2$, we thus necessarily have $n=11$ or $n\ge 13$.
Moreover, we also get
\begin{equation}\label{eq:3}
|\overline{S}|=n-\left\lfloor\frac{n+2}{2}\right\rfloor \ge 3q - n + 2\left\lfloor\frac{n+2}{2}\right\rfloor - 2 \ge 3q-1 \ge q+3 \ge 5.
\end{equation}

The edge colouring $\gamma_1$ is then obtained as follows, depending on the parity of $q$.

\begin{enumerate}

\item $q=2t$, $t\ge 1$.\\
We choose any set of $t$ pairs of vertices $X=\{(x_i,y_i)\}_{1\le i\le t}$ in $\overline{S}$.
%(this is possible since, by~\eqref{eq:3}, $|\overline{S}|\ge q+4$).
Since $ux_ivy_i$ is a 2-monochromatic 4-cycle for every $i$, $1\le i\le t$, we can
apply the recolouring of type~2 to the set $X$, so that
$|E_{\gamma_1}(1)|=|E_{\gamma_1}(3)|=|E_{\gamma_0}(1)|+q$
and $|E_{\gamma_1}(2)|=|E_{\gamma_0}(2)|-2q$.
The so-obtained edge colouring $\gamma_1$ is thus
an equitable good  edge 3-nsd-colouring of $K_{n+2}$.

\medskip

\item $q=2t+1$, $t\ge 1$.\\
We first choose any set of $t$ pairs of vertices $X=\{(x_i,y_i)\}_{1\le i\le t-1}$
in $\overline{S}\setminus\{w_{min}^{\gamma},w_{max}^{\gamma}\}$
(this is possible since, by~\eqref{eq:3}, $|\overline{S}|\ge q+3$).
Since $ux_ivy_i$ is a 2-monochromatic 4-cycle for every $i$, $1\le i\le t$, we can
apply the recolouring of type~2 to the set $X$, so that
we have $|E_{\gamma_0}(1)|+q-1$ edges coloured by 1 (resp. by 3).
We then apply the recolouring of type~1, since $uw_{min}^{\gamma}vw_{max}^{\gamma}$
is still a 2-monochromatic 4-cycle,
%The set of vertices $\{u,v,x,y,z\}$ still induces a 2-monochromatic $K_5$
%and we can thus apply the recolouring of type~3 to this set,
so that
$|E_{\gamma_1}(1)|=|E_{\gamma_1}(3)|=|E_{\gamma_0}(1)|+q$
and $|E_{\gamma_1}(2)|=|E_{\gamma_0}(2)|-2q$.
%Since $|\overline{S}|\ge q+4$ by~\eqref{eq:3},
%there exists a vertex $w\in\overline{S}\setminus\{x,y,z,x_1,y_1,\dots,x_{t-1},y_{t-1}\}$
%and the triangle $uvw$ is 2-monochromatic.

Therefore, the so-obtained edge colouring $\gamma_1$ is
an equitable good  edge 3-nsd-colouring of $K_{n+2}$.

\end{enumerate}

This concludes the proof of Theorem~\ref{TheoremEquitableSumCompleteGraph}.
\qed
\end{pf}

%%%%%%%%%% XXXXXXXXXXXXX
\section{Proof of Theorem~\ref{TheoremEquitableSumCompleteBipartiteGraph}}

We prove in this section that $\EchiE(K_{m,n})=2$ whenever $m=n=2$ or $m=n\geq 4$,
$\EchiE(K_{3,3})=3$ and $\EchiE(K_{m,n})=1$ if $1\leq m<n$.
As in the previous section, we denote by $E_{\gamma}(i)$ the set of edges
that are assigned colour $i$ by the edge colouring $\gamma$.

\begin{pf}[of Theorem~\ref{TheoremEquitableSumCompleteBipartiteGraph}]
If $1\le m<n$, then adjacent vertices have distinct degrees, and hence colouring all
edges with colour 1 gives an equitable  edge 1-nsd-colouring of $K_{m,n}$.

\medskip

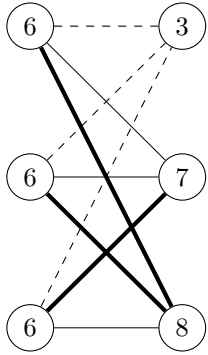
\begin{figure}[t]
\begin{center}
\begin{tikzpicture}

\def \n {4}
\def \radius {1cm}
\def \hradius {2cm}
\def \nradius {0.5cm}
\def \margin {8} % margin in angles, depends on the radius
\node (r0) at ({-\radius},\hradius) [draw, circle,fill=none, minimum size=\nradius] {$6$};
\node (r1) at ({-\radius},0) [draw, circle,fill=none, minimum size=\nradius] {$6$};
\node (r2) at ({-\radius},{-\hradius}) [draw, circle,fill=none, minimum size=\nradius] {$6$};
\node (r3) at (\radius, {\hradius}) [draw, circle,fill=none, minimum size=\nradius] {$3$};
\node (r4) at (\radius,0) [draw, circle,fill=none, minimum size=\nradius] {$7$};
\node (r5) at (\radius,{-\hradius}) [draw, circle,fill=none, minimum size=\nradius] {$8$};

\draw [dashed] (r0) -- (r3);
\draw [ultra thin] (r0) -- (r4);
\draw [ultra thick] (r0) -- (r5);
\draw [dashed] (r1) -- (r3);
\draw [ultra thin] (r1) -- (r4);
\draw [ultra thick] (r1) -- (r5);
\draw [dashed] (r2) -- (r3);
\draw [ultra thick] (r2) -- (r4);
\draw [ultra thin] (r2) -- (r5);

\end{tikzpicture}
\caption{\label{fig:K33Color} An equitable  edge 3-nsd-colouring of $K_{3,3}$.}
\end{center}
\end{figure}

If $m=n$, then $K_{m,n}$ is regular, and therefore, $\overline{\chi_{\Sigma}^e}(K_{m,n}) \ge 2$.
Suppose first that $m=n=3$ and let $V\cup V'$ denotes the bipartition of $V(K_{3,3})$,
with $V=\{v_1,v_2,v_3\}$ and $V'=\{v'_1,v'_2,v'_3\}$.
We first claim that $\overline{\chi_{\Sigma}^e}(K_{3,3}) > 2$. Assume to the contrary
that $\gamma$ is an equitable  edge 2-nsd-colouring of $K_{3,3}$
and let $\sigma$ denote the vertex colouring induced by $\gamma$.
Since $|E(K_{3,3})|=9$, we necessarily have
\begin{equation}\label{eq:4or5}
\{|E_{\gamma}(1)|,|E_{\gamma}(2)|\}=\{4,5\}.
\end{equation}

Moreover, since $3\le \sigma(v)\le 6$ for every vertex $v\in V\cup V'$, we get
without loss of generality either $\sigma(v_1)=\sigma(v_2)=\sigma(v_3)$,
or $\sigma(v_1)=\sigma(v_2)\neq\sigma(v'_1)=\sigma(v'_2)$.
In the first case, we get $|E_{\gamma}(1)| + 2|E_{\gamma}(2)| = 3\sigma(v_1)\equiv 0\pmod 3$, in contradiction with~\eqref{eq:4or5}.
In the latter case, we necessarily have $\{\sigma(v_1),\sigma(v'_1)\}=\{4,5\}$, since otherwise
we would have six edges with the same colour.
Assume without loss of generality that the edges incident with $v_1$ and $v_2$ are coloured 1, 1 and 2.
Since $\gamma$ is an equitable edge colouring, the edges incident with $v_3$ are necessarily coloured 2, 2, and 2, but then $\sigma(v'_3)=4=\sigma(v_1)$, or 2, 2 and 1, again
a contradiction since this would in turn imply $\sigma(v_3)=\sigma(v'_1)=5$.
Taking into account the equitable  edge 3-nsd-colouring of $K_{3,3}$
depicted in Figure~\ref{fig:K33Color}, we get $\overline{\chi_{\Sigma}^e}(K_{3,3})=3$.

\medskip

Finally, suppose that $m=n=2$ or $m=n\ge 4$ and let $V\cup V'$ denote the bipartition of $V(K_{n,n})$,
with $V=\{v_1,\dots,v_n\}$ and $V'=\{v'_1,\dots,v'_n\}$.
We consider two cases, depending on the parity of $n$.

\begin{enumerate}
\item $n=2t$, $t\ge 1$.\\
Let $\gamma$ be the edge 2-colouring of $K_{n,n}$ defined as follows.
For every edge $v_iv'_j\in E(K_{n,n})$, let $\sigma(v_iv'_j)=1$ if $i$ is odd
and $\sigma(v_iv'_j)=2$ otherwise. Since $n$ is even, $\gamma$ is an equitable
edge 2-colouring. To see that $\gamma$ is neighbour-sum-distinguishing, observe
that for every $i$, $1\le i\le n$, $\sigma(v_i)=2t$ if $i$ is odd,
$\sigma(v_i)=4t$ if $i$ is even,
while $\sigma(v'_j)=3t$ for every $j$, $1\le j\le n$.
\medskip

\item $n=2t+1$, $t\ge 2$.\\
Let $\gamma$ be the edge 2-colouring of $K_{n,n}$ defined as follows.
For the subgraph of $K_{n,n}$ induced by $\{v_1,\dots,v_{n-1}\}\cup\{v'_1,\dots,v'_{n-1}\}$,
$\gamma$ is defined as in the previous case.
We then set
$\gamma(v_nv'_j)=1$ for every $j$, $1\le j\le n-1$,
$\gamma(v_iv'_n)=2$ for every $i$, $1\le i\le n-1$,
and $\sigma(v_nv'_n)=1$.
The edge colouring $\gamma$ thus obtained is clearly an equitable edge 2-colouring.
To see that $\gamma$ is neighbour-sum-distinguishing, observe that for every $i$, $1\le i\le n-1$,
$\sigma(v_i)=2t+2$ if $i$ is odd,
$\sigma(v_i)=4t+2$ if $i$ is even,
while $\sigma(v'_j)=3t+1$ for every $j$, $1\le j\le n-1$,
$\sigma(v_n)=2t+1$ and $\sigma(v'_n)=4t+1$.
\end{enumerate}

This concludes the proof of Theorem~\ref{TheoremEquitableSumCompleteBipartiteGraph}.
\qed\end{pf}

\section{Proof of Theorem~\ref{TheoremEquitableSumForest}}
%\begin{pf}

%- If there is an induced path $v_0,v_1,\ldots,v_k$ in a graph $G$ such that $v_1,\ldots,v_k$ have no neighbours in $G$ outside this path (i.e. $d_G(v_k)=1$, $d_G(v_{k-1})=2,\ldots,d_G(v_1)=2$), we call this path a \emph{pendant path} %(\emph{at} $v_0$)
%of length $k$ incident with $v_0$. In particular, if $k=1$, we call $v_0v_1$ a \emph{pendant edge}.\\ %appended?
%- Also in a graph (in a forest?) $G$, a vertex of degree $1$ will be called a \emph{leaf} or a \emph{pendant vertex}.\\

We prove in this section that $\EchiE(F)\le 2$ for every forest $F$ with no isolated edge.
Throughout the proof we will thus use only colours $1$ and $2$ to colour the edges.

A vertex of degree~1 will be called a \emph{leaf} or a \emph{pendant vertex}.
If $v_0,v_1,\ldots,v_k$ is an induced path such that $v_1,\ldots,v_k$ have no neighbours in $G$ outside this path
(i.e. $d_G(v_1)=\ldots=d_G(v_{k-1})=2$ and $d_G(v_k)=1$), we call this path a \emph{pendant path} %(\emph{at} $v_0$)
of length $k$ incident with $v_0$.
In particular, if $k=1$, we call $v_0v_1$ a \emph{pendant edge}.

Suppose a forest $F$ is a \emph{minimal counterexample}, i.e., a counterexample with minimal number of edges, to Theorem~\ref{TheoremEquitableSumForest}.

\begin{claim}\label{ClaimNoPath}
No component of $F$ is a path.
\end{claim}

\begin{pf}
Suppose $P$ is a component of $F$ which is a path.
Then we colour the forest $F'$ obtained of $F$ by removing all vertices of $P$ \emph{by the minimality of $F$},
what will mean here and in all further claims that we fix some  equitable (edge) nsd-colouring of $F'$ with $1$ and $2$, which exists due to the fact that $F$ is a minimal counterexample to Theorem~\ref{TheoremEquitableSumForest} (in cases where we will be left with components $K_2$ in $F'$, what does not take place in this claim, we will mean that we colour the forest formed by the remaining components of $F'$ by the minimality of $F$ and then we put $1$'s or $2$'s on the isolated edges of $F'$ so that the colouring is equitable).

%(i.e., by the fact that $F$ is a counterexample )
Now it is sufficient to colour the path $P$ equitably so that its neighbours are sum-distinguished and the colouring of entire $F$ is equitable
in order to obtain an equitable nsd-colouring of $F$, a contradiction with the fact that $F$ is a
counterexample to Theorem~\ref{TheoremEquitableSumForest}.
In case when $P$ is of even length it suffices to use the same number of $1$'s and $2$'s, while for odd path $P$ we might be forced to use one
more $1$ or $2$ (and we do not control which one).
As leaves are always sum-distinguished from their neighbours,
this can however be always easily achieved, as we only need to colour every second edge of the path differently, i.e., it is always sufficient to colour appropriately the first two edges of the path -- the rest of the colours on the path are the consequence of these two
(note also here a fact useful in further reasonings that if in a graph $G$ we have a pendant path of length $4$, then its edges must be coloured with two $1$'s and two $2$'s in any equitable  edge nsd-colouring of $G$ with $1$ and $2$).
\qed
\end{pf}

Since we want to prove that no counterexample to Theorem~\ref{TheoremEquitableSumForest} exists,
i.e., that in fact $\EchiE(F)\le 2$, we may make use of the following reduction.
%Suppose first that $T$ is a path (with more than one edge). As leaves are always sum-distinguished from their neighbours,
%in order to obtain a sum-distinguishing colouring of a path we must use a different weight on its every second edge.
%This can be always done equitably - it is sufficient to start with $1$ and $2$ on the first two edges of $T$.
%
%Suppose now that $T$ is a counterexample to the thesis (hence it cannot be a path) with minimum number of edges $m$ ($m\geq 3$).
%
%We make the following simplifying observation.
\begin{claim}\label{ClaimReduction}
We may assume that $F$ contains no vertex $u$ of degree $3$ adjacent to a leaf $w$ and a vertex $x$ of degree $2$ whose other neighbour $y$ is a leaf.
\end{claim}
\begin{pf}
Suppose there is such a vertex $u$, and let $v$ be its remaining neighbour in $F$ ($v\notin \{w,x\}$), see Figure~\ref{forests_figure_1}(a).
%%%%%%%%%%%%%%%%%%%%%%%%%%%%%%%%%%%%%%%%%%%%%%%%%%%%%%%%%%%%%%%%%%%%-fig1
%+++++++++++++++++++++++++++++++++++++++++++++++++++++++++++++++++++
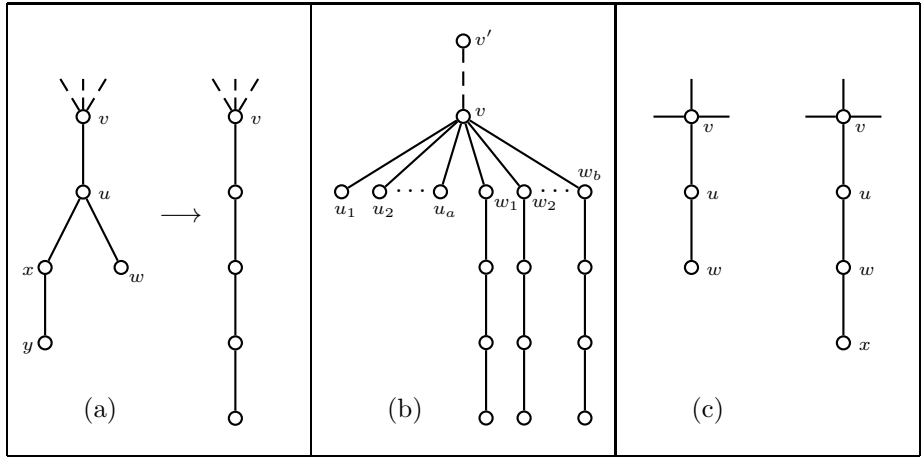
\begin{figure}%\label{konfi}
\setlength{\unitlength}{1cm}
%\psset{unit=1cm}
\psset{radius=0.1}
\begin{center}
\begin{picture}(12,6)

%rama-proste poziome
\multiput(0,0)(0,6){2}{\line(1,0){12}}
%\put(0,0){\line(1,0){130}}
%\put(0,40){\line(1,0){130}}
%\put(0,120){\line(1,0){130}}

%rama-proste pionove
\multiput(0,0)(4,0){4}{\line(0,1){6}}
% \put(20,0){\line(0,1){120}}
%\put(80,0){\line(0,1){120}}
%\put(130,0){\line(0,1){120}}
%%%%%%%%%%%%%%%%%%------p1
\put(1,3.5){\begin{pspicture}(2,1)%------------------------------------------left,left
%\psset{unit=0.7cm}

\Cnode(0,1){c}
\Cnode(0,0){a0}
\Cnode(-0.5,-1){a1}
\Cnode(0.5,-1){b1}
\Cnode(-0.5,-2){a2}
\pnode(-0.3,1.5){c1}
\pnode(0,1.5){c2}
\pnode(0.3,1.5){c3}
\ncline{a0}{a1}
\ncline{a2}{a1}
\ncline{a0}{b1}
\ncline{a0}{c}
\ncline[linestyle=dashed]{c1}{c}
\ncline[linestyle=dashed]{c2}{c}
\ncline[linestyle=dashed]{c3}{c}
\put(0.2,-0.1){{\scriptsize $u$}}
\put(0.2,0.9){{\scriptsize $v$}}
\put(-0.8,-1.1){\scriptsize $x$}
\put(-0.8,-2.1){\scriptsize $y$}
\put(0.6,-1.2){\scriptsize $w$}
\end{pspicture}
}
%%%%%%%%%%%%%%%%%%%%%%%%%%%%%%%%%%%%%%%%
\put(3,3.5){\begin{pspicture}(2,1)%------------------------------------------left,right
%\psset{unit=0.7cm}

\Cnode(0,1){c}
\Cnode(0,0){a0}
\Cnode(0,-1){a1}
\Cnode(0,-3){a3}
\Cnode(0,-2){a2}
\pnode(-0.3,1.5){c1}
\pnode(0,1.5){c2}
\pnode(0.3,1.5){c3}
\ncline{a0}{a1}
\ncline{a2}{a1}
\ncline{a2}{a3}
\ncline{a0}{c}
\ncline[linestyle=dashed]{c1}{c}
\ncline[linestyle=dashed]{c2}{c}
\ncline[linestyle=dashed]{c3}{c}
%\put(0.2,-0.1){{\scriptsize $u$}}
\put(0.2,0.9){{\scriptsize $v$}}
%\put(-0.8,-1.1){\scriptsize $x$}
%\put(-0.8,-2.1){\scriptsize $y$}
%\put(0.6,-1.2){\scriptsize $w$}
\end{pspicture}
}
%%%%%%%%%%%%%%%%%%%%%%%%%%%%%%%%%%%%%%%
\put(2,3.1){$\longrightarrow$}

%%%%%%%%%%%%%%%%%%%%%%%%%%%%%%%%%%%%%%%%%%%%
\put(6,4.5){\begin{pspicture}(1,1)%-------------------------middle
%\psset{unit=0.7cm}

\Cnode(0,1){c}
\Cnode(0,0){a0}
\Cnode(-1.6,-1){a1}
\Cnode(-1.1,-1){a2}
\Cnode(-0.3,-1){a3}
\put(-0.9,-1){$\ldots$}

\Cnode(1.6,-1){b1}
\Cnode(0.8,-1){b2}
\Cnode(0.3,-1){b3}
\put(1,-1){$\ldots$}
\Cnode(1.6,-2){x1}
\Cnode(1.6,-3){x2}
\Cnode(1.6,-4){x3}

\Cnode(0.8,-2){y1}
\Cnode(0.8,-3){y2}
\Cnode(0.8,-4){y3}

\Cnode(0.3,-2){z1}
\Cnode(0.3,-3){z2}
\Cnode(0.3,-4){z3}

%\pnode(-0.3,1.5){c1}
%\pnode(0,1.5){c2}
%\pnode(0.3,1.5){c3}
\ncline{a0}{a1}
\ncline{a0}{a2}
\ncline{a0}{a3}
\ncline{a0}{b1}
\ncline{a0}{b2}
\ncline{a0}{b3}
\ncline[linestyle=dashed]{a0}{c}
\ncline{b1}{x1}
\ncline{x1}{x2}
\ncline{x2}{x3}

\ncline{b2}{y1}
\ncline{y1}{y2}
\ncline{y2}{y3}

\ncline{b3}{z1}
\ncline{z1}{z2}
\ncline{z2}{z3}
%\ncline[linestyle=dashed]{c1}{c}
%\ncline[linestyle=dashed]{c2}{c}
%\ncline[linestyle=dashed]{c3}{c}
\put(0.15,0){{\scriptsize $v$}}
\put(0.15,0.95){{\scriptsize $v'$}}
\put(-1.7,-1.3){\scriptsize $u_1$}
\put(-1.2,-1.3){\scriptsize $u_2$}
\put(-0.4,-1.3){\scriptsize $u_a$}

\put(0.4,-1.2){\scriptsize $w_1$}
\put(0.9,-1.2){\scriptsize $w_2$}
\put(1.5,-0.8){\scriptsize $w_b$}
\end{pspicture}
}
%%%%%%%%%%%%%%%%%%%%%%%%%%%%%%%%%%%%%%%%%%%%%%%%%%%%%%%%%%%%%%%%%
%%%%%%%%%%%%%%%%%%%%%%%%%%%%%%%%%%%%%%%%
\put(9,3.5){\begin{pspicture}(2,1)%------------------------------------------right,left
%\psset{unit=0.7cm}

\Cnode(0,1){c}
\Cnode(0,0){a0}
\Cnode(0,-1){a1}
%\Cnode(0,-2){a2}
\pnode(-0.5,1){c1}
\pnode(0,1.5){c2}
\pnode(0.5,1){c3}
\ncline{a0}{a1}
%\ncline{a2}{a1}
\ncline{a0}{c}
\ncline{c1}{c}
\ncline{c2}{c}
\ncline{c3}{c}
\put(0.2,-0.1){{\scriptsize $u$}}
\put(0.15,0.8){{\scriptsize $v$}}
%\put(-0.8,-1.1){\scriptsize $x$}
\put(0.2,-1.1){\scriptsize $w$}
\end{pspicture}
}

%%%%%%%%%%%%%%%%%%%%%%%%%%%%%%%%%%%%%%%%
\put(11,3.5){\begin{pspicture}(2,1)%------------------------------------------right,right
%\psset{unit=0.7cm}

\Cnode(0,1){c}
\Cnode(0,0){a0}
\Cnode(0,-1){a1}
\Cnode(0,-2){a2}
\pnode(-0.5,1){c1}
\pnode(0,1.5){c2}
\pnode(0.5,1){c3}
\ncline{a0}{a1}
\ncline{a2}{a1}
\ncline{a0}{c}
\ncline{c1}{c}
\ncline{c2}{c}
\ncline{c3}{c}
\put(0.2,-0.1){{\scriptsize $u$}}
\put(0.15,0.8){{\scriptsize $v$}}
\put(0.2,-2.1){\scriptsize $x$}
\put(0.2,-1.1){\scriptsize $w$}
\end{pspicture}
}

\put(1,0.51){(a)}
\put(5,0.51){(b)}
\put(9,0.51){(c)}

%%%%%%%%%%%%%%%%%%%%%%%%%%%%%%%%%%%%%%%%%%%%%%%%%%%%%%%%%%%%%%%%%

\end{picture}

\caption{Illustrations to Claims~\ref{ClaimReduction}, \ref{Claim1and4paths} and \ref{Claim23pathsDeg4+}.}
\label{forests_figure_1}
\end{center}
\end{figure}

%++++++++++++++++++++++++++
%%%%%%%%%%%%%%%%%%%%%%%%%%%%%%%%%%%%%%%%%%%%%%%%%%%%%%%%%%%%%%%%%%%%%%%%%%-fig1-end
Then $\overline{\chi_{\Sigma}^e}(F)\leq 2$ only if $\overline{\chi_{\Sigma}^e}(F')\leq 2$ where $F'$ is the forest obtained from $F$ by deleting the vertices $u,w,x,y$
(together with their four incident edges) and appending a pendant path of length $4$ at $v$, i.e., identifying one end of this path with $v$ (so that the numbers of edges in $F$ and $F'$ are equal).
Suppose that there is an equitable edge colouring of $F'$ with $1$ and $2$ (obviously, $\overline{\chi_{\Sigma}^e}(F')\neq 1$).
Then use the same colouring on all edges of $F$ that appear also in $F'$, so that four edges of $F$ remain uncoloured.
Note that in order to be certain that the colouring of $F$ is equitable we must use colours $1,1,2,2$ on these four remaining edges, as exactly these four colours must have been used on the pendant path of length $4$ in $F'$ (cf. the argument for paths above).
First we copy on $uv$ the colour from the edge incident with $v$ in $F'$ from the mentioned path of length $4$
%which was incident with $v$ in $T'$
(in order to avoid sum conflicts between $v$ and its neighbours other than $u$).
Then we put a colour on $uw$ so that $uv$ and $uw$ have distinct colours, hence we are left with $1$ and $2$ to use.
We choose one of these colours for $ux$ so that there is no conflict between $u$ and $v$, and we use the remaining colour on $xy$.
Note that by our construction the sum at $x$ will always be smaller than the sum at $u$.

Hence, as $F$ is a minimal (i.e., with minimum number of edges) counterexample to Theorem~\ref{TheoremEquitableSumForest}, then so does $F'$.
We may thus perform the operation described above repeatedly until there are no configurations from the thesis in our forest left.
\qed
\end{pf}

Let us root every tree (component) of $F$ at any leaf. A vertex $v$ of degree at least $3$ with all descendants of degree at most $2$ will be called a \emph{last multifather} (this is just a vertex which induces with its descendants only pendant paths incident with this vertex).
First we present a few observations implying that %for any vertex $v$ of degree at least $3$, if its all descendants have degree at most $2$, then in fact they all have to be leaves
all descendants of any last multifather must in fact be leaves
(in other words, all pendant paths incident with such a vertex and containing its descendants are in fact pendant edges),
see Claim~\ref{ClaimLastMultifathers} below.
Some of these observations will be  also useful in the further part of the argument, e.g. the following seemingly very specific claim.

\begin{claim}\label{Claim1and4paths}
There is no vertex $v$ of degree at most $a+b+1$ incident
with $a$ ($a\geq 0$) pendant edges and $b$ ($b\geq 1$) pendant paths of length $4$ in $F$.
%The forest $F$ contains no vertex $v$ of degree at most $a+b+1$ %adjacent
%with $a\geq 0$ pendant edges and $b\geq 1$ pendant paths of length $4$ at it.
\end{claim}

\begin{pf}
Suppose to the contrary that  $u_1,\ldots,u_a$ are $a$ ($a\geq 0$) leaves adjacent with $v$, while $w_1,\ldots,w_b$ 
are $b$ ($b\geq 1$) neighbours of $v$ such that $vw_i$ is the first of four edges of a pendant path of length $4$ incident with $v$, $i=1,\ldots,b$, see Figure~\ref{forests_figure_1}(b). Denote by $v'$ the remaining neighbour of $v$ (if there is any). %Let $T'$ be a subtree of $G$ being the component of $T-\{u_1,\ldots,u_a,w_1,\ldots,w_b\}$ containing $v$.

Let $F'$ be the forest obtained of $F$ by deleting $u_1,\ldots,u_a$ and all $4b$ vertices (except $v$) from the $b$ pendant paths of length $4$ incident with $v$ (including $w_1,\ldots,w_b$, resp.).
By the minimality of $F$, $F'$ admits an equitable 2-colouring.
%By the minimality of $F$, $T'$ cannot be a counterexample to the theorem, so may colour it as desired. (Note that $T'$ might also be an isolated vertex or edge. In the second case, here and in similar circumstances below, we will colour it e.g. with $1$).
It suffices then to complete the colouring using an appropriate equitable number of $1$'s and $2$'s.

If $a=0$, we may use the same number of $1$'s and $2$'s. Otherwise, we first greedily choose colours for $vu_1,\ldots,vu_a$ so that %we are left with equal number of $1$'s and $2$'s (i.e., $2b$ of each).
we obtain a partial equitable colouring of $F$ (i.e., the number of $1$' and $2$'s used so far on $F$ is as equal as possible).
We will then use $2b$ $1$'s and $2b$ $2$'s on the remaining edges.
For this goal we first choose any %of the remaining
colours for $vw_1,\ldots,vw_b$ so that there is no conflict between $v$ and $v'$.
Finally, for each of the pendant paths of length $4$ incident with $v$, we complete its colouring (similarly as in the case of a path itself above) first using a colour on yet non-coloured edge incident with $w_i$ to avoid conflict between $w_i$ and $v$, and so on, in order to obtain an nsd-colouring of $F$ with $1$ and $2$.
Note that on each such path we will use two $1$'s and two $2$'s, thus the colouring will also be equitable, a contradiction with the minimality of $F$. \qed
\end{pf}

As a consequence of the claim above, for $a=0$ and $b=1$, we obtain the following:

\begin{claim}\label{Claim5+paths}
There are no pendant paths of length (at least) $5$ in $F$. %A path of length at least $5$ cannot be pendant to any vertex of $T$.
\end{claim}

We supplement this observation with the two following ones.

\begin{claim}\label{Claim23pathsDeg4+}
There is no vertex $v$ with $\deg(v)\geq 4$ incident with a pendant path of length $2$ or $3$ in $F$.
%In $F$, there is no vertex $v$ with a pendant path of length $2$ or $3$ at it and $\deg(v)\geq 4$.
%in
%A path of length $2$ or $3$ cannot be pendant to a vertex $v$ with $\deg(v)\geq 4$.
\end{claim}

\begin{pf}
Suppose there is such a vertex $v$ in $F$, and let $v,u,w$ or $v,u,w,x$ be the consecutive vertices of the corresponding path, see Figure~\ref{forests_figure_1}(c).
By the minimality of $F$, we may colour $F-\{uw\}$ or $F-\{uw,wx\}$, respectively. In the first case, we conclude by using on $uw$
any of the available at most $2$ colours -- note that as $\deg(v)\geq 4$ and $\deg(u)=2$, the sum at $v$ will always be greater than the one at $u$.
In the second case, we use $1$ or $2$ on $wx$ so that there is no conflict between $u$ and $w$, 
%and put a different colour on $uw$, 
and colour the edge $uv$ in such a way that the colouring is equitable,
a contradiction.
\qed
\end{pf}

\begin{claim}\label{Claim2pathsDeg3}
A vertex $v$ of degree $3$ cannot be incident with both a pendant path of length at least 1 and a pendant
path of length at least 2 in $F$.
%two pendant paths of length at least $1$ one of which has length at least $2$ in $F$.
%Two paths of lengths at least $1$, one of which has length at lest $2$ cannot be pendant to a vertex $v$ of degree $3$ in $T$.
\end{claim}

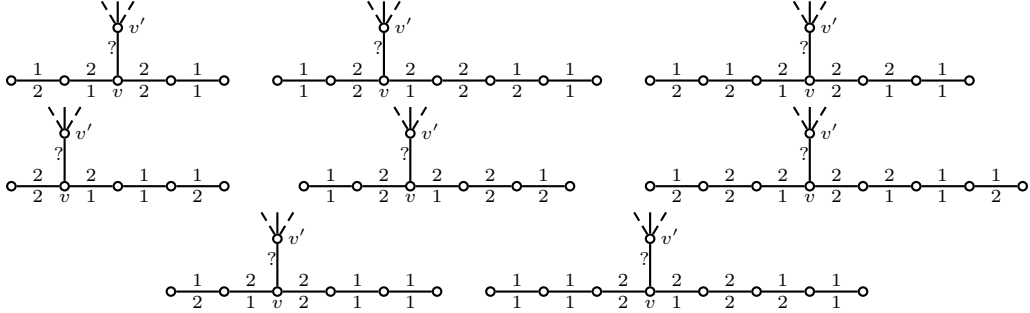
\begin{figure}%\label{konfi}
\psset{unit=0.7cm}
\psset{radius=0.1}
%\begin{center}
\begin{pspicture}(13,6)

%*************************************************

%\Cnode[fillstyle=solid,fillcolor=gray](3,5){A0}{6}
%\Cnode[fillstyle=solid,fillcolor=black](6,5){A1}
%\Cnode(9,5){A2}

%\psset{linewidth=3mm}
%\rput(4,0.5){\circlenode{A3}{\scriptsize{$\alpha$}}}
%\rput(5,0.5){\circlenode{A4}{\tiny{$\beta$}}}

%***************************************************************************************
\put(2,4.5){\begin{pspicture}(2,1)%------------------------------------------p11
%\psset{unit=0.7cm}

\Cnode(0,1){c}
\Cnode(0,0){a0}
\Cnode(1,0){a1}
\Cnode(2,0){a2}
\Cnode(-1,0){b1}
\Cnode(-2,0){b2}
\pnode(-0.3,1.5){c1}
\pnode(0,1.5){c2}
\pnode(0.3,1.5){c3}
\ncline{a0}{a1}
\ncline{a2}{a1}
\ncline{a0}{b1}
\ncline{b2}{b1}
\ncline{a0}{c}
\ncline[linestyle=dashed]{c1}{c}
\ncline[linestyle=dashed]{c2}{c}
\ncline[linestyle=dashed]{c3}{c}
\put(-0.1,-0.3){{\scriptsize $v$}}
\put(0.2,0.9){{\scriptsize $v'$}}
\put(-0.2,0.5){{\scriptsize ?}}
\put(0.4,0.1){\scriptsize $2$}
\put(0.4,-0.3){\scriptsize $2$}
\put(1.4,0.1){\scriptsize $1$}
\put(1.4,-0.3){\scriptsize $1$}
\put(-0.6,0.1){\scriptsize $2$}
\put(-0.6,-0.3){\scriptsize $1$}
\put(-1.6,0.1){\scriptsize $1$}
\put(-1.6,-0.3){\scriptsize $2$}
\end{pspicture}
}
%***************************************************************************************
\put(7,4.5){\begin{pspicture}(2,1)%------------------------------------------p12
%\psset{unit=0.7cm}

\Cnode(0,1){c}
\Cnode(0,0){a0}
\Cnode(1,0){a1}
\Cnode(2,0){a2}
\Cnode(3,0){a3}
\Cnode(4,0){a4}
\Cnode(-1,0){b1}
\Cnode(-2,0){b2}
\pnode(-0.3,1.5){c1}
\pnode(0,1.5){c2}
\pnode(0.3,1.5){c3}
\ncline{a0}{a1}
\ncline{a2}{a1}
\ncline{a2}{a3}
\ncline{a3}{a4}
\ncline{a0}{b1}
\ncline{b2}{b1}
\ncline{a0}{c}
\ncline[linestyle=dashed]{c1}{c}
\ncline[linestyle=dashed]{c2}{c}
\ncline[linestyle=dashed]{c3}{c}
\put(-0.1,-0.3){{\scriptsize $v$}}
\put(0.2,0.9){{\scriptsize $v'$}}
\put(-0.2,0.5){{\scriptsize ?}}
\put(0.4,0.1){\scriptsize $2$}
\put(0.4,-0.3){\scriptsize $1$}
\put(1.4,0.1){\scriptsize $2$}
\put(1.4,-0.3){\scriptsize $2$}
\put(2.4,0.1){\scriptsize $1$}
\put(2.4,-0.3){\scriptsize $2$}
\put(3.4,0.1){\scriptsize $1$}
\put(3.4,-0.3){\scriptsize $1$}
\put(-0.6,0.1){\scriptsize $2$}
\put(-0.6,-0.3){\scriptsize $2$}
\put(-1.6,0.1){\scriptsize $1$}
\put(-1.6,-0.3){\scriptsize $1$}
\end{pspicture}
}
%---------------------------------------
\put(15,4.5){\begin{pspicture}(2,1)%------------------------------------------p13
%\psset{unit=0.7cm}

\Cnode(0,1){c}
\Cnode(0,0){a0}
\Cnode(1,0){a1}
\Cnode(2,0){a2}
\Cnode(3,0){a3}
\Cnode(-1,0){b1}
\Cnode(-2,0){b2}
\Cnode(-3,0){b3}
\pnode(-0.3,1.5){c1}
\pnode(0,1.5){c2}
\pnode(0.3,1.5){c3}
\ncline{a0}{a1}
\ncline{a2}{a1}
\ncline{a2}{a3}
\ncline{a0}{b1}
\ncline{b2}{b1}
\ncline{b2}{b3}
\ncline{a0}{c}
\ncline[linestyle=dashed]{c1}{c}
\ncline[linestyle=dashed]{c2}{c}
\ncline[linestyle=dashed]{c3}{c}
\put(-0.1,-0.3){{\scriptsize $v$}}
\put(0.2,0.9){{\scriptsize $v'$}}
\put(-0.2,0.5){{\scriptsize ?}}
\put(0.4,0.1){\scriptsize $2$}
\put(0.4,-0.3){\scriptsize $2$}
\put(1.4,0.1){\scriptsize $2$}
\put(1.4,-0.3){\scriptsize $1$}
\put(2.4,0.1){\scriptsize $1$}
\put(2.4,-0.3){\scriptsize $1$}
\put(-0.6,0.1){\scriptsize $2$}
\put(-0.6,-0.3){\scriptsize $1$}
\put(-1.6,0.1){\scriptsize $1$}
\put(-1.6,-0.3){\scriptsize $2$}
\put(-2.6,0.1){\scriptsize $1$}
\put(-2.6,-0.3){\scriptsize $2$}
\end{pspicture}
}
%---------------------------------------
%***************************************************************************************
\put(1,2.5){\begin{pspicture}(2,1)%------------------------------------------p21
%\psset{unit=0.7cm}

\Cnode(0,1){c}
\Cnode(0,0){a0}
\Cnode(1,0){a1}
\Cnode(2,0){a2}
\Cnode(3,0){a3}
%\Cnode(4,0){a4}
\Cnode(-1,0){b1}
%\Cnode(-2,0){b2}
\pnode(-0.3,1.5){c1}
\pnode(0,1.5){c2}
\pnode(0.3,1.5){c3}
\ncline{a0}{a1}
\ncline{a2}{a1}
\ncline{a2}{a3}
%\ncline{a3}{a4}
\ncline{a0}{b1}
%\ncline{b2}{b1}
\ncline{a0}{c}
\ncline[linestyle=dashed]{c1}{c}
\ncline[linestyle=dashed]{c2}{c}
\ncline[linestyle=dashed]{c3}{c}
\put(-0.1,-0.3){{\scriptsize $v$}}
\put(0.2,0.9){{\scriptsize $v'$}}
\put(-0.2,0.5){{\scriptsize ?}}
\put(0.4,0.1){\scriptsize $2$}
\put(0.4,-0.3){\scriptsize $1$}
\put(1.4,0.1){\scriptsize $1$}
\put(1.4,-0.3){\scriptsize $1$}
\put(2.4,0.1){\scriptsize $1$}
\put(2.4,-0.3){\scriptsize $2$}
%\put(3.4,0.1){\scriptsize $1$}
%\put(3.4,-0.3){\scriptsize $2$}
\put(-0.6,0.1){\scriptsize $2$}
\put(-0.6,-0.3){\scriptsize $2$}
%\put(-1.6,0.1){\scriptsize $1$}
%\put(-1.6,-0.3){\scriptsize $2$}
\end{pspicture}
}
%***************************************************************************************
\put(7.5,2.5){\begin{pspicture}(2,1)%------------------------------------------p22
%\psset{unit=0.7cm}

\Cnode(0,1){c}
\Cnode(0,0){a0}
\Cnode(1,0){a1}
\Cnode(2,0){a2}
\Cnode(3,0){a3}
%\Cnode(4,0){a4}
\Cnode(-1,0){b1}
\Cnode(-2,0){b2}
\pnode(-0.3,1.5){c1}
\pnode(0,1.5){c2}
\pnode(0.3,1.5){c3}
\ncline{a0}{a1}
\ncline{a2}{a1}
\ncline{a2}{a3}
%\ncline{a3}{a4}
\ncline{a0}{b1}
\ncline{b2}{b1}
\ncline{a0}{c}
\ncline[linestyle=dashed]{c1}{c}
\ncline[linestyle=dashed]{c2}{c}
\ncline[linestyle=dashed]{c3}{c}
\put(-0.1,-0.3){{\scriptsize $v$}}
\put(0.2,0.9){{\scriptsize $v'$}}
\put(-0.2,0.5){{\scriptsize ?}}
\put(0.4,0.1){\scriptsize $2$}
\put(0.4,-0.3){\scriptsize $1$}
\put(1.4,0.1){\scriptsize $2$}
\put(1.4,-0.3){\scriptsize $2$}
\put(2.4,0.1){\scriptsize $1$}
\put(2.4,-0.3){\scriptsize $2$}
%\put(3.4,0.1){\scriptsize $1$}
%\put(3.4,-0.3){\scriptsize $1$}
\put(-0.6,0.1){\scriptsize $2$}
\put(-0.6,-0.3){\scriptsize $2$}
\put(-1.6,0.1){\scriptsize $1$}
\put(-1.6,-0.3){\scriptsize $1$}
\end{pspicture}
}
%---------------------------------------
\put(15,2.5){\begin{pspicture}(2,1)%------------------------------------------p23
%\psset{unit=0.7cm}

\Cnode(0,1){c}
\Cnode(0,0){a0}
\Cnode(1,0){a1}
\Cnode(2,0){a2}
\Cnode(3,0){a3}
\Cnode(4,0){a4}
\Cnode(-1,0){b1}
\Cnode(-2,0){b2}
\Cnode(-3,0){b3}
\pnode(-0.3,1.5){c1}
\pnode(0,1.5){c2}
\pnode(0.3,1.5){c3}
\ncline{a0}{a1}
\ncline{a2}{a1}
\ncline{a2}{a3}
\ncline{a4}{a3}
\ncline{a0}{b1}
\ncline{b2}{b1}
\ncline{b2}{b3}
\ncline{a0}{c}
\ncline[linestyle=dashed]{c1}{c}
\ncline[linestyle=dashed]{c2}{c}
\ncline[linestyle=dashed]{c3}{c}
\put(-0.1,-0.3){{\scriptsize $v$}}
\put(0.2,0.9){{\scriptsize $v'$}}
\put(-0.2,0.5){{\scriptsize ?}}
\put(0.4,0.1){\scriptsize $2$}
\put(0.4,-0.3){\scriptsize $2$}
\put(1.4,0.1){\scriptsize $2$}
\put(1.4,-0.3){\scriptsize $1$}
\put(2.4,0.1){\scriptsize $1$}
\put(2.4,-0.3){\scriptsize $1$}
\put(3.4,0.1){\scriptsize $1$}
\put(3.4,-0.3){\scriptsize $2$}
\put(-0.6,0.1){\scriptsize $2$}
\put(-0.6,-0.3){\scriptsize $1$}
\put(-1.6,0.1){\scriptsize $2$}
\put(-1.6,-0.3){\scriptsize $2$}
\put(-2.6,0.1){\scriptsize $1$}
\put(-2.6,-0.3){\scriptsize $2$}
\end{pspicture}
}
%---------------------------------------
%***************************************************************************************
%***************************************************************************************
\put(5,0.5){\begin{pspicture}(2,1)%------------------------------------------p32
%\psset{unit=0.7cm}

\Cnode(0,1){c}
\Cnode(0,0){a0}
\Cnode(1,0){a1}
\Cnode(2,0){a2}
\Cnode(3,0){a3}
%\Cnode(4,0){a4}
\Cnode(-1,0){b1}
\Cnode(-2,0){b2}
\pnode(-0.3,1.5){c1}
\pnode(0,1.5){c2}
\pnode(0.3,1.5){c3}
\ncline{a0}{a1}
\ncline{a2}{a1}
\ncline{a2}{a3}
%\ncline{a3}{a4}
\ncline{a0}{b1}
\ncline{b2}{b1}
\ncline{a0}{c}
\ncline[linestyle=dashed]{c1}{c}
\ncline[linestyle=dashed]{c2}{c}
\ncline[linestyle=dashed]{c3}{c}
\put(-0.1,-0.3){{\scriptsize $v$}}
\put(0.2,0.9){{\scriptsize $v'$}}
\put(-0.2,0.5){{\scriptsize ?}}
\put(0.4,0.1){\scriptsize $2$}
\put(0.4,-0.3){\scriptsize $2$}
\put(1.4,0.1){\scriptsize $1$}
\put(1.4,-0.3){\scriptsize $1$}
\put(2.4,0.1){\scriptsize $1$}
\put(2.4,-0.3){\scriptsize $1$}
%\put(3.4,0.1){\scriptsize $1$}
%\put(3.4,-0.3){\scriptsize $1$}
\put(-0.6,0.1){\scriptsize $2$}
\put(-0.6,-0.3){\scriptsize $1$}
\put(-1.6,0.1){\scriptsize $1$}
\put(-1.6,-0.3){\scriptsize $2$}
\end{pspicture}
}
%---------------------------------------
\put(12,0.5){\begin{pspicture}(2,1)%------------------------------------------p33
%\psset{unit=0.7cm}

\Cnode(0,1){c}
\Cnode(0,0){a0}
\Cnode(1,0){a1}
\Cnode(2,0){a2}
\Cnode(3,0){a3}
\Cnode(4,0){a4}
\Cnode(-1,0){b1}
\Cnode(-2,0){b2}
\Cnode(-3,0){b3}
\pnode(-0.3,1.5){c1}
\pnode(0,1.5){c2}
\pnode(0.3,1.5){c3}
\ncline{a0}{a1}
\ncline{a2}{a1}
\ncline{a2}{a3}
\ncline{a4}{a3}
\ncline{a0}{b1}
\ncline{b2}{b1}
\ncline{b2}{b3}
\ncline{a0}{c}
\ncline[linestyle=dashed]{c1}{c}
\ncline[linestyle=dashed]{c2}{c}
\ncline[linestyle=dashed]{c3}{c}
\put(-0.1,-0.3){{\scriptsize $v$}}
\put(0.2,0.9){{\scriptsize $v'$}}
\put(-0.2,0.5){{\scriptsize ?}}
\put(0.4,0.1){\scriptsize $2$}
\put(0.4,-0.3){\scriptsize $1$}
\put(1.4,0.1){\scriptsize $2$}
\put(1.4,-0.3){\scriptsize $2$}
\put(2.4,0.1){\scriptsize $1$}
\put(2.4,-0.3){\scriptsize $2$}
\put(3.4,0.1){\scriptsize $1$}
\put(3.4,-0.3){\scriptsize $1$}
\put(-0.6,0.1){\scriptsize $2$}
\put(-0.6,-0.3){\scriptsize $2$}
\put(-1.6,0.1){\scriptsize $1$}
\put(-1.6,-0.3){\scriptsize $1$}
\put(-2.6,0.1){\scriptsize $1$}
\put(-2.6,-0.3){\scriptsize $1$}
\end{pspicture}
}
%---------------------------------------
\end{pspicture}

%\rput(2,1){\circlenode[fillstyle=solid, fillcolor=lightgray, linestyle=none]{X1}{\scriptsize $1$}}
%\end{pspicture}

%\end{center}
\caption{Subcases of Claim~\ref{Claim2pathsDeg3}.}
\label{forests_figure_2}
\end{figure}

\begin{pf}
Suppose there is such a vertex $v$ in $F$. Note that by Claim~\ref{Claim5+paths}, the both paths have to be of length at most $4$,
and one of them has to have length at most $3$ by Claim~\ref{Claim1and4paths} (with $a=0$ and $b=2$).
Additionally, if one of the paths is just a pendant edge, then the other cannot have length $2$, by Claim~\ref{ClaimReduction}, nor $4$, by Claim~\ref{Claim1and4paths}.

We thus are left with $6$ cases.
In each of these cases, we first colour by the minimality of $F$ the forest $F'$ obtained from $F$ by removing all edges of the two pendant paths.
If there is an even number of such edges, we then use an even number of $1$'s and $2$'s to complete the colouring.
Otherwise we might be forced to use one more $1$ or $2$ so that the colouring of $F$ is equitable at the end.
Thus, for each such case, we analyse the two corresponding subcases.
For each of these cases (and subcases) we start by choosing the colours for the first edges of the two paths (those incident with $v$) appropriately so that $v$ is not in conflict with its neighbour $v'$ from $F'$.
In  Figure~\ref{forests_figure_2} we  show that in all (sub)cases there are always two possible choices (with different sums, one of which must be appropriate) for these two edges, which then can be extended without conflicts, regardless of the colour of $vv'$ (marked thus by ``$?$''), on the remaining yet uncoloured edges (one choice is presented above the edges, while the alternative is presented below).
We thus obtain a contradiction with the minimality of $F$ as a counterexample to Theorem~\ref{TheoremEquitableSumForest}.
\qed
\end{pf}

%%%%%%%%%%%%%%%%%%%%%%%%%%%%%%%%%%%%%%%%%%%%%%%%%%%%%%%%%%%%%%%%%%%%%%%%%%%%%%%%%%%%-fig2
%%%%%%%%%%%%%%%%%%%%%%%%%%%%%%%%%%%%%%%%%%%%%%%%%%%%%%%%%%%%%%%%%%%%%%%%%%%%%%%%%%%%%
%%%%%%%%%%%%%%%%%%%%%%%%%%%%%%%%%%%%%%%%%%%%%%%%%%%%%%%%%%%%%%%%%%%%%%%%%%%
%%%%%%%%%%%%%%%%%%%%%%%%%%%%%%%%%%%%%%%%%%%%%%%%%%%%%%%%%%%%%%%%%%%%%%%%%%%%%%%%%%%%%-fig2-end

\begin{claim}\label{ClaimLastMultifathers}
Every vertex $v$ of degree at least $3$ with all descendants of degree at most $2$ in $F$ has only descendants of degree $1$.
\end{claim}

\begin{pf}
For vertices of degree at least $4$ this follows by Claims~\ref{Claim5+paths}, \ref{Claim23pathsDeg4+} and~\ref{Claim1and4paths}, while for vertices of degree $3$ this is a consequence of Claim~\ref{Claim2pathsDeg3}.
\qed
\end{pf}

\begin{claim}\label{ClaimFatherOFLastMultifather}
Every vertex $v$ of degree at least $3$ in $F$ with all descendants of degree $1$ has a father of degree at least $3$.
\end{claim}

\begin{pf}
Suppose there is a vertex $v$ in $F$ of degree at least $3$ with descendants being leaves and a father $v'$ of degree at most $2$.
Then we delete all edges joining $v$ with its descendants in $F$ and colour the obtained forest by the minimality of $F$.
To conclude it is sufficient to colour the pendant edges incident with $v$ using at least one $2$ (this will be possible, and even almost always necessary) so that the colouring of $F$ is equitable. This way $v$ has certainly a greater sum than $v'$, a contradiction.
\qed
\end{pf}

Note that by Claims~\ref{ClaimNoPath},~\ref{ClaimLastMultifathers} and~\ref{ClaimFatherOFLastMultifather}, every component of $F$ must in particular have at least two vertices of degree greater than $3$.
Thus each such component $T$ (previously rooted at some leaf) must contain at least one vertex which we will call a \emph{last multigrandfather},
that is a vertex $v$ of degree at least $3$ in $T$ which has at least one descendant of degree at least $3$ but none of the descendants of degree at least $3$ of $v$ has further descendants of degree at least $3$
(i.e., all descendants of degree at least $3$ of $v$ are last multifathers).
Note that such a $v$ is adjacent with its father while, due to Claims~\ref{ClaimLastMultifathers}, \ref{ClaimFatherOFLastMultifather} and~\ref{Claim5+paths},
 %its descendants (and $v$) form %stars or paths of lengths at most $4$ pendant at $v$
%pendant paths of lengths at most $4$ incident with $v$
%(i.e.,
every son of $v$ either has degree at least $3$ and all sons being leaves, or has degree $2$ and at most $3$ descendants - all of degree at most $2$. %).
Moreover, by Claim~\ref{Claim23pathsDeg4+}, if $\deg(v)\geq 4$, every %such
pendant path incident with $v$ must have length exactly $1$ or $4$.

Below, see Claim~\ref{ClaimNoMultigrandfathers}, we obtain a contradiction with the statement above that every component of $F$ contains a last multigrandfather.
Consequently, we will prove that no counterexample to Theorem~\ref{TheoremEquitableSumForest} may exist, thus concluding its proof.
%In the remaining part of the argument we show that in all cases, we can first delete from $T$ some descendants of arbitrarily chosen last %multigrandfather in order to use the minimality of $F$ and then complete (and possibly slightly change) the equitable colouring of $T$ so that %neighbours are sum-distinguished. As $T$ must contain at least one last multigrandfather, we will thus obtain a contradiction, proving the fact %that no counterexample to Theorem~\ref{TheoremEquitableSumForest} exists.

\begin{claim}\label{ClaimNoMultigrandfathers}
No component $T$ of $F$ %The tree $T$
contains %no
a last multigrandfather.
\end{claim}

\begin{pf}
Assume to the contrary that $v$ is a last multigrandfather in a component $T$ of~$F$.

Suppose first that $v$ has two sons $u$ and $w$ of degree at least $3$.
Then we may delete two pendant edges incident with $u$, say $e_u,e'_u$ and two pendant edges incident with $w$, say $e_w,e'_w$ (recall that all descendants of $u$ and $w$ must be leaves),
and colour the remaining forest by the minimality of $F$.
It is then sufficient to colour the remaining four edges with two $1$'s and two $2$'s so that there is no conflict between $v$ and its sons $u,w$ in order to get a contradiction with the minimality of $F$.
We however have three essentially different ways of extending our colouring, i.e., assigning $1,1$ to $e_u,e'_u$ (and hence $2,2$ to $e_w,e'_w$) or assigning $1,2$ or $2,2$ to them. Hence, one of these options must fulfil our requirements (as the sum at $v$ ``forbids'' only one potential sum at each of $u$ and $w$).

%As $\deg(v)\geq 3$ we are left with...
Hence we may assume that $v$ has exactly one son who is a last multifather, say $u$ (if it had no such son, it could not be a last multigrandfather by definition).
Denote by $u',u''$ any two leaves adjacent with $u$.

Let us consider first the case when all the remaining sons of $v$ are leaves.
If $\deg(v)\geq 4$, then delete two pendant edges incident with $u$, say $e_u,e'_u$ and two pendant edges incident with $v$, say $e_v,e'_v$,
and colour the remaining forest by the minimality of $F$.
Then, analogously as above, it is sufficient to colour the remaining four edges with two $1$'s and two $2$'s so that there is no conflict between $u$ and $v$ and between $v$ and its father.
As we have three essentially different ways of extending the colouring,  at most two of which being ``forbidden'' by our requirements on lack of conflicts, we certainly may extend the colouring to $F$.
If however $\deg(v)=3$, let $w$ be the son of $v$ which is a leaf. %adjacent with $v$.  %and let $u',u''$ be any two leaves adjacent with $u$.
Delete from $F$ all edges induced by $v$ and its descendants and colour the remaining forest by the minimality of $F$.
As we have removed at least four edges, we certainly may use at least two $1$'s and at least two $2$'s while equitably extending the colouring to $F$.
Assign $1$ to $vw$ and $2$ to $uu'$.
Then choose $\gamma(uv)\in\{1,2\}$, where $\gamma(uv)$ denotes the colour of the edge $uv$, so that there is no conflict between $v$ and its father $v'$,
 set $\gamma(uu'')=3-\gamma(uv)$, and assign colours to the remaining uncoloured edges (if there are any) so that the obtained colouring of $F$ is equitable.
 If there is no conflict between $u$ and $v$, we are done.
On the other hand, the only situation in which we may have such a conflict is when $u',u''$ are the only sons of $u$, $\gamma(vv')=2$,
%where $v'$ is the father of $v$ and $\gamma(vv')$ denotes the colour of the edge $vv'$,
and $\gamma(uu'')=1$ (in all other cases the sum at $u$ would be larger than the one at $v$), and hence $\gamma(uv)=2$, see Figure~\ref{forests_figure_3}(a).
%%%%%%%%%%%%%%%%%%%%%%%%%%%%%%%%%%%%%%%%%%%%%%%%%%%%%%%%%%%%%%%%%%%%%%%%%%%%%%-fig3
%+++++++++++++++++++++++++++++++++++++++++++++++++++++++++++++++++++
\begin{figure}%\label{konfi}
\setlength{\unitlength}{1cm}
%\psset{unit=1cm}
\psset{radius=0.1}
\begin{center}
\begin{picture}(13,6)

%rama-proste poziome
\multiput(0,0)(0,6){2}{\line(1,0){13}}
%\put(0,0){\line(1,0){130}}
%\put(0,40){\line(1,0){130}}
%\put(0,120){\line(1,0){130}}

%rama-proste pionove
\multiput(0,0)(2.5,0){2}{\line(0,1){6}}
\put(8.5,0){\line(0,1){6}}
%\put(80,0){\line(0,1){120}}
\put(13,0){\line(0,1){6}}
%%%%%%%%%%%%%%%%%%------p1
\put(1,3.5){\begin{pspicture}(2,1)%\showgrid%(2,1)%------------------------left
%\psset{unit=0.7cm}

\Cnode(0,1){c}
\Cnode(0,0){a0}
\Cnode(-0.5,-1){a1}
\Cnode(0.5,-1){b1}
\Cnode(0,-2){b2}
\Cnode(1,-2){b3}
%\Cnode(-0.5,-2){a2}
\pnode(-0.3,1.5){c1}
\pnode(0,1.5){c2}
\pnode(0.3,1.5){c3}
\ncline{a0}{a1}
%\ncline{a2}{a1}
\ncline{a0}{b1}
\ncline{b2}{b1}
\ncline{b3}{b1}
\ncline{a0}{c}
\ncline[linestyle=dashed]{c1}{c}
\ncline[linestyle=dashed]{c2}{c}
\ncline[linestyle=dashed]{c3}{c}
\put(0.2,-0.1){{\scriptsize $v$}}
\put(0.2,0.9){{\scriptsize $v'$}}
\put(-0.8,-1.2){\scriptsize $w$}
\put(0.65,-1){\scriptsize $u$}
\put(0,-2.25){\scriptsize $u'$}
\put(1,-2.25){\scriptsize $u''$}
\put(0.1,0.5){{\scriptsize $2$}}
\put(0.17,-0.8){{\scriptsize $2$}}
%\put(-0.2,-0.8){\red $\leftrightarrow$}
\psline[linecolor=red]{<->}(0.17,-0.7)(-0.17,-0.7)
\put(-0.32,-0.8){\scriptsize $1$}
%\put(-0.7,-1.6){\scriptsize $1$}
\put(0.3,-1.6){\scriptsize $2$}
\put(0.9,-1.6){\scriptsize $1$}
\end{pspicture}
}
%%%%%%%%%%%%%%%%%%%%%%%%%%%%%%%%%%%%%%%%
\put(4,3.5){\begin{pspicture}(2,1)%------------------------------------------middle,left
%\psset{unit=0.7cm}

\Cnode(0,1){c}
\Cnode(0,0){a0}
\Cnode(-0.5,-1){a1}
\Cnode(-1,-2){a2}
\Cnode(0.5,-1){b1}
\Cnode(0,-2){b2}
\Cnode(0.5,-2){b3}
\Cnode(1.2,-2){b4}
\put(0.65,-2){$\ldots$}
\pnode(-0.3,1.5){c1}
\pnode(0,1.5){c2}
\pnode(0.3,1.5){c3}
\ncline{a0}{a1}
\ncline{a2}{a1}
\ncline{a0}{b1}
\ncline{b2}{b1}
\ncline{b3}{b1}
\ncline[linestyle=dashed]{b1}{b4}
\ncline{a0}{c}
\ncline[linestyle=dashed]{c1}{c}
\ncline[linestyle=dashed]{c2}{c}
\ncline[linestyle=dashed]{c3}{c}
\put(0.2,-0.1){{\scriptsize $v$}}
\put(0.2,0.9){{\scriptsize $v'$}}
\put(-0.8,-1.2){\scriptsize $w$}
\put(-1,-2.25){\scriptsize $w'$}
\put(0.65,-1){\scriptsize $u$}
\put(0,-2.25){\scriptsize $u'$}
\put(0.5,-2.25){\scriptsize $u''$}
%----------------------------------------------
\put(0.1,0.5){{\scriptsize ?}}
\put(0.4,-0.7){{\scriptsize ?}}
\put(-0.3,-0.8){\scriptsize $1$}
\put(-0.7,-1.6){\scriptsize $1$}
\put(0.3,-1.8){\scriptsize $2$}
\put(0,-1.6){\scriptsize $2$}
\end{pspicture}
}
%%%%%%%%%%%%%%%%%%%%%%%%%%%%%%%%%%%%%%%
\put(5,3.5){\blue $\longrightarrow$}
%%%%%%%%%%%%%%%%%%%%%%%%%%%%%%%%%%%%%%%%%%%%
\put(7,3.5){\begin{pspicture}(2,1)%------------------------------------------middle,right
%\psset{unit=0.7cm}

\Cnode(0,1){c}
\Cnode(0,0){a0}
\Cnode(-0.5,-1){a1}
\Cnode(-1,-2){a2}
\Cnode(0.5,-1){b1}
\Cnode(0,-2){b2}
\Cnode(1,-2){b3}
\pnode(-0.3,1.5){c1}
\pnode(0,1.5){c2}
\pnode(0.3,1.5){c3}
\ncline{a0}{a1}
\ncline{a2}{a1}
\ncline{a0}{b1}
\ncline{b2}{b1}
\ncline{b3}{b1}
\ncline{a0}{c}
\ncline[linestyle=dashed]{c1}{c}
\ncline[linestyle=dashed]{c2}{c}
\ncline[linestyle=dashed]{c3}{c}
\put(0.2,-0.1){{\scriptsize $v$}}
\put(0.2,0.9){{\scriptsize $v'$}}
\put(-0.8,-1.2){\scriptsize $w$}
\put(-1,-2.25){\scriptsize $w'$}
\put(0.65,-1){\scriptsize $u$}
\put(0,-2.25){\scriptsize $u'$}
\put(1,-2.25){\scriptsize $u''$}
%----------------------------------------------
\put(0.1,0.5){{\scriptsize ?}}
\put(0.4,-0.7){{\scriptsize $2$}}
\put(-0.3,-0.8){\scriptsize $2$}
\put(-0.7,-1.6){\scriptsize ?}
\put(0.3,-1.6){\scriptsize $1$}
\put(0.9,-1.6){\scriptsize $1$}
\end{pspicture}
}
%%%%%%%%%%%%%%%%%%%%%%%%%%%%%%%%%%%%%%%
\put(11,4.25){\begin{pspicture}(2,1)%\showgrid%--------------------------right
%\psset{unit=0.7cm}

\Cnode(0,1){c}
\Cnode(0,0){a0}
\Cnode(-1,-1){a1}
\Cnode(-1,-2){a2}
\Cnode(-1,-3){y1}
\Cnode(-1,-4){y2}
\Cnode(-2,-1){x1}
\Cnode(-2,-2){x2}
\Cnode(-2,-3){x3}
\Cnode(-2,-4){x4}
\put(-1.7,-3){$\ldots$}
\Cnode(0.5,-1){b1}
\Cnode(0,-2){b2}
\Cnode(0.5,-2){b3}
\Cnode(1.4,-2){b4}
\put(0.7,-2){$\ldots$}
\pnode(-0.3,1.5){c1}
\pnode(0,1.5){c2}
\pnode(0.3,1.5){c3}
\ncline{a0}{a1}
\ncline{a2}{a1}
\ncline{a2}{y1}
\ncline[linestyle=dashed]{y2}{y1}
\ncline{a0}{b1}
\ncline{b2}{b1}
\ncline{b3}{b1}
\ncline[linestyle=dashed]{b4}{b1}
\ncline{a0}{c}
\ncline[linestyle=dashed]{a0}{x1}
\ncline[linestyle=dashed]{x2}{x1}
\ncline[linestyle=dashed]{x2}{x3}
\ncline[linestyle=dashed]{x4}{x3}
\ncline[linestyle=dashed]{c1}{c}
\ncline[linestyle=dashed]{c2}{c}
\ncline[linestyle=dashed]{c3}{c}
\put(0.2,-0.1){{\scriptsize $v$}}
\put(0.2,0.9){{\scriptsize $v'$}}
\put(-1.4,-1.2){\scriptsize $w_1$}
\put(-1.4,-2.2){\scriptsize $w_2$}
\put(-0.9,-3.1){\scriptsize $w_3$}
\put(-0.9,-4.1){\scriptsize $w_4$}
\put(0.65,-1){\scriptsize $u$}
\put(0,-2.25){\scriptsize $u'$}
\put(0.5,-2.25){\scriptsize $u''$}
%----------------------------------------------
\put(0.1,0.5){{\scriptsize ?}}
\put(0.1,-0.7){{\scriptsize $1$}}
\put(-0.5,-0.7){\scriptsize $1$}
\put(-1.2,-1.6){\scriptsize $1$}
\put(-0.9,-2.6){\scriptsize $2$}
\put(-0.9,-3.6){\green \scriptsize $2$}
\psline[linecolor=red]{<->}(0.15,-0.75)(0.15,-1.3)
%\put(0.1,-1){\red \scriptsize $\updownarrow$}
\psline[linecolor=red]{<->}(-0.45,-0.75)(-0.8,-2.4)
\put(0.1,-1.5){\scriptsize $2$}
\put(0.5,-1.6){\scriptsize $?$}
\end{pspicture}
}
%%%%%%%%%%%%%%%%%%%%%%%%%%%%%%%%%%%%%%%

\put(1.1,0.5){(a)}
\put(5.5,0.5){(b)}
\put(12,0.5){(c)}

%%%%%%%%%%%%%%%%%%%%%%%%%%%%%%%%%%%%%%%%%%%%%%%%%%%%%%%%%%%%%%%%%

\end{picture}

\caption{Illustrations to cases in Claim~\ref{ClaimNoMultigrandfathers}.}
\label{forests_figure_3}
\end{center}
\end{figure}
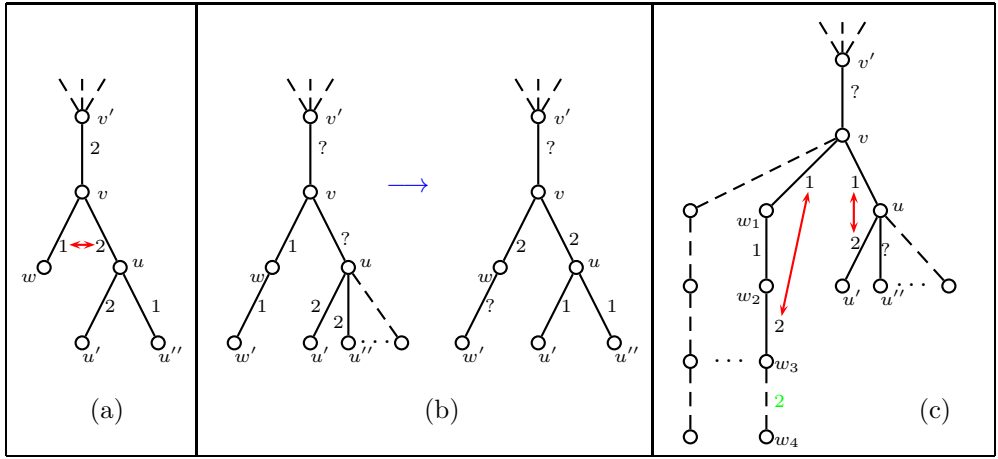
%%%%%%%%%%%%%%%%%%%%%%%%%%%%%%%%%%%%%%%%%%%%%%%%%%%%%%%%%%%%%%%%%%%%%%%%%%%%%%%%%%%-fig3-end
But then we may switch the colours of $uv$ and $vw$, decreasing the sum at $u$ but not changing the sum at $v$.
In all cases we thus obtain a desired  equitable nsd-colouring of $F$, a contradiction.

Suppose now that $v$ has a son, say $w$, adjacent with only one leaf $w'$ (i.e., $v,w,w'$ form a pendant path of length $2$ incident with $v$).
By Claim~\ref{Claim23pathsDeg4+}, $\deg(v)=3$, see Figure~\ref{forests_figure_3}(b).
Then delete all descendants of $v$ and colour the forest obtained by the minimality of $F$.
As we have removed at least five edges from $F$, we still may use at least two $1$'s and two $2$'s.
Set $\gamma(ww')=1=\gamma(wv)$, $\gamma(uu')=2=\gamma(uu'')$ and if we still have any choice (which will not prevent us from %being able to
completing the colouring of $F$ equitably) choose a colour for $uv$ so that there is no conflict between $v$ and its father.
Finally, colour the remaining uncoloured edges equitably (i.e. so that the colouring of $F$ is equitable).
This way, no conflict is possible, except a potential conflict between $v$ and its father.
In such a case, we must however have had no choice while colouring $uv$, and hence $\deg(u)=3$.
Then we colour all previously removed edges once more differently, setting $\gamma(vw)=2=\gamma(vu)$ (and hence increasing the sum at $v$),
$\gamma(uu')=1=\gamma(uu'')$, and completing the colouring equitably, see Figure~\ref{forests_figure_3}(b).
As no conflict is then possible, we obtain a contradiction.

Suppose finally that there is a pendant path of length $3$ or $4$ incident with $v$ and induced by $v$ and its descendants.
Delete the edges of this path and all edges incident with $u$ and colour the obtained forest by the minimality of $F$.
Let $w_1,w_2,\ldots,w_j$, $j\in \{3,4\}$ be the consecutive vertices of this path with $w_1$ being a son of $v$.
Set $\gamma(vw_1)=1=\gamma(w_1w_2)$, $\gamma(w_2w_3)=2$ and $\gamma(w_3w_4)=2$ (if there is such an edge), $\gamma(vu)=1$, $\gamma(uu')=2$, see Figure~\ref{forests_figure_3}(c),
and colour the remaining edges so that we obtain an equitable colouring of $F$.
Note that as $\gamma(w_1w_2)=1$ (what will not be changed), no conflict is possible between $w_1$ and $v$.
We may however have potential conflicts between $v$ %and $u$ and between $v$ and its father $v'$.
and its remaining neighbours.
If there is a conflict between $v$ and $u$ or $v'$, where $v'$ is the father of $v$, %one of this conflicts arises,
we exchange the colours of $vu$ and $uu'$, increasing the sum at $v$ (and not changing the sum at $u$ nor at $v'$).
After such a switch, since hitherto there was a conflict, $v$ must have a greater sum than $u$ or $v'$.
If $v$ is still in conflict with the remaining one of these two, we raise the sum at $v$ once more (not changing the sums at $u$ and $v'$) by switching the colours of $vw_1$ and $w_2w_3$.
Then if there was still some conflict in $F$, it would have to be between $v$ and its neighbour, say $x$, other than $v',u$ and $w_1$,
but then $\deg(v)\geq 4$ and $\deg(x)\leq 2$ (as we have assumed that $v$ has only one son of degree at least $3$), hence $v$ and $x$ could not be in conflict.
%(and %by... every path pendant at $v$ has length $1$ or $4$).
%But then this $x_1$ must belong to some path $v,x_1,x_2,x_3,x_4$ of length $4$ pendant at $v$.
%Then however, the edges of this

Thus, in all cases we have been able to obtain a desired  equitable nsd-colouring of $F$ using colours $1$ and $2$, a contradiction.
\qed
\end{pf}

%This completes the proof of Theorem~\ref{TheoremEquitableSumForest}.
%\qed
%\end{pf}

\section{Proof of Theorem~\ref{TheoremEquitableSumTotalBipartite}}

%ZZZ

We prove in this section that  $\EchiT(G)\le 2$ for every bipartite graph $G$.
In order to prove the result in the non-connected case, we will in fact prove a stronger thesis but in the case of connected bipartite graphs.
We will not only prove that such graphs admit  equitable total nsd-colourings using $1$ and $2$,
but also that, if the sum of the numbers of vertices and edges is odd, then there are two such colourings -- one with a majority of $1$'s and the second with a majority of $2$'s.
This immediately implies the thesis of Theorem~\ref{TheoremEquitableSumTotalBipartite} (as we may first colour the components of a non-connected bipartite graph with even sums of numbers of vertices and edges, and then the remaining ones, using alternately a majority of $1$'s and a majority of $2$'s).

Note that this strengthened thesis for connected bipartite graphs is straightforward in the case of a star (even with no edges) -- e.g., if a star has at least two edges, it is sufficient to put $1$'s on all its edges and $2$'s on the vertices or the other way round.
%The strengthened thesis for
For the remaining connected bipartite graphs it follows %also immediately
by first using Observation~\ref{ObsMajorityOf1Bipartite}, and then sequentially repeating application of Lemma~\ref{LemmaBipartite2Exchanging1} below until we achieve one or two desired total colourings.

\begin{observation}\label{ObsMajorityOf1Bipartite}
Every connected bipartite graph $G=(X,Y;E)$ with at least one edge can be totally coloured with $1$ and $2$ so that the vertices in one set of the bipartition have even sums and the vertices in the second set of the bipartition have odd sums and so that the number of $1$'s used exceeds the non-zero number of $2$'s.
\end{observation}

\begin{pf}
Colour all the edges of $G$ with $1$. Then colour one vertex in $X$ with $2$. Next subsequently colour all the remaining vertices in $G$, each with $1$ or $2$, so that the parities of sums at all the vertices in $X$ are the same and different from those in $Y$.
If the number of $1$'s used on $G$ does not exceed the number of $2$'s, it means that all vertices are coloured with $2$ (note that the number of vertices may exceed the number of edges by at most one in a connected graph).
Then choose any edge $uv\in E$ and change the colours of its end-vertices from $2$ to $1$ and the colour of the edge from $1$ to $2$.
Note that this will not influence the sums at any vertex in $G$, but the number of $1$'s used will exceed the number of $2$'s afterwards
(while at least one $2$ will remain on $G$).
\qed
%each with $1$ or $2$ so that the sums at the vertices in $X$ are even and odd in $Y$. If the number of $1$'s used on $G$ does not exceed the number %of $2$'s,
%change the colours of all vertices in $G$. Then certainly the number of $1$'s used shell exceed the number of $2$'s, while the sums at the vertices %in $X$ will become odd and those in $Y$ - even. \qed
\end{pf}

Note that neighbours are certainly sum-distinguished under the colouring from Observation~\ref{ObsMajorityOf1Bipartite} above.
Now we will show that given such a colouring we can repeatedly increase the number of $2$'s used (at each step only by one) not spoiling at the same time the neighbour distinction in $G$ until we achieve our goal (or goals).

\begin{lemma}\label{LemmaBipartite2Exchanging1}
Given any total colouring with $1$ and $2$ of a connected bipartite graph $G=(V,E)$ which is not a star and
with the number of $1$'s used exceeding the non-zero number of $2$'s
such that the vertices in one set of the bipartition of $G$ have even sums and the vertices in the second set of the bipartition have odd sums,
we may construct a new total colouring of $G$ with $1$ and $2$ complying with the second feature of the given one (concerning the parities of the sums at vertices) but with exactly one more $2$ used than in the initial colouring.
\end{lemma}

\begin{pf}
Suppose we are given a graph and an initial colouring as claimed.
The proof will be based on the fact that the parities of the sums in $G$ do not change if we \emph{make a negative} of any edge $uv$, %(or a few edges one after another),
i.e., after changing every $1$ to $2$ and every $2$ to $1$ used on $u$, $v$ and $uv$.
%We will write that we tak a negative of a colour/1...

We will show that starting from our initial colouring we may always subsequently make negatives of a few edges to obtain a colouring consistent with the thesis.
We will write that an edge $uv$ is \emph{of type} $abc$, where $a,b,c\in\{1,2\}$ if $u,uv,v$ (or $v,uv,u$) are coloured $a,b,c$, respectively, in the initial colouring.
Note first that
%we may assume that (initially) there is no edge $uv$ in $G$ coloured $2$ with $u$ and $v$ coloured $1$ nor $uv$ coloured $1$ with %one end coloured $1$ and the other $2$, as we could make a negative of such an edge end immediately obtain our goal.
%Thus
we may assume that there are no edges of types 121 and 112 (nor 211) in $G$, as we could make a negative of any such edge and immediately achieve our goal.

Suppose there is no edge of type 111 in $G$ either, hence there are only edges of types 222, 212 and $221$ (or equivalently $122$).
%Then however we can show that we would have more $2$'s than $1$'s used on $G$ (a contradiction with the assumptions on our initial colouring). %Indeed
Then the graph $H$ induced in $G$ by the edges of type $212$ cannot be a forest, as otherwise there would be more $2$'s than $1$'s on $G$.
Indeed, if $H$ is a forest, i.e., has more vertices than edges, then more $2$'s appear on vertices than $1$'s on edges in $G$.
On the other hand, there are at least as many edges coloured $2$ as there are vertices coloured $1$ in $G$, as every such vertex must be an end of an edge of type $221$ (or $122$), and since the other end of such an edge is coloured $2$,
 we may easily define an injective mapping from the set of vertices coloured $1$ to the set of edges coloured $2$ (by assigning to such a vertex any of its incident edges).
Hence, as $H$ is not a forest and must be bipartite, it contains a cycle of length at least $4$, and hence also a path of length $3$.
It is then sufficient to subsequently make negatives of all these three edges (after which the consecutive edges of this path of types 212, 212, 212 will become edges of types 122, 222, 221 respectively -- note in particular that the vertices of the middle edge will switch colours twice, hence in fact will return to their initial values) to obtain a required colouring of $G$.

Suppose then to the contrary that there is an edge of type $111$ in $G$.
As the parities of the sums at the ends of such an edge $e$ must differ (by the definition of our initial colouring),
$e$ must be adjacent with at least one edge coloured $1$, say $f$, thus $f$ is also an edge of type $111$ (as there are no edges of type 112 nor 211 in $G$).
Note that we may then assume that there is no edge of type $222$, as otherwise it could not be adjacent with any edge of type $111$ and thus we could make negatives of such an edge of type 222 and two adjacent edges of type 111, and obtain a required colouring of $G$.
Thus in $G$ there are only edges of type 111, which we will call edges of type $A$, and edges of types $212$ and $122$ (or $221$), which we will all call edges of type $B$.
Analogously as above, one may verify that if there is a path of length $3$ in $G$ with two consecutive edges of type $A$ and one of type $B$ or two consecutive edges of type $B$ and one of type $A$, then by making consecutively negatives of all edges of one such path, we will always obtain a desired total colouring of $G$.
We will show that such a path must exist in $G$.
Let $H'$ be any component of a graph induced in $G$ by the edges of type $A$. %with the largest diameter
(Recall that each such component must have diameter at least $2$.)

Suppose that %diameter is at least $3$.
$H'$ is not a star.
As $G$ is connected and at least one $2$ is used as a colour on it, at least one vertex, say $v$, (coloured $1$) in $H'$ must be incident with an edge, say $e'$ of type $B$.
Note that the other (different from $v$) end of $e'$, say $u$, must be coloured $2$, hence does not belong to $H'$.
On the other hand, in $H'$ there must be a vertex, say $w$, at distance $2$ from $v$, as otherwise $H'$ would be a star.
Hence there is a path of length $3$ (starting at $u$ and ending at $w$) in $G$ with two consecutive edges of type $A$ and one of type $B$, as claimed.

We thus may finally assume that $H'$ is a star (with at least two edges). If at least one of its leaves is incident with an edge of type $B$, then we obtain a path as above.
Otherwise however, as $G$ is connected, $G$ is not a star and at least one $2$ was used on it, the center of the star making up $H'$ must be incident with one end of a path $P$ of length two, whose first edge (incident with the center of the star), say $e'$, is of type $B$.
However, as the other end of $e'$ must be coloured with $2$, the second edge of this path must also be of type $B$, thus we obtain a path of length $3$ with two consecutive edges of type $B$ (from $P$) and one edge of type $A$ (incident with the center of $H'$), as claimed.
\qed
\end{pf}

\section{Proof of Theorem~\ref{TheoremEquitableSumTotalComplete}}

Finally, we prove in this section that  $\EchiT(K_2)=2$ and $\EchiT(K_n)=3$ for every $n\ge 3$.

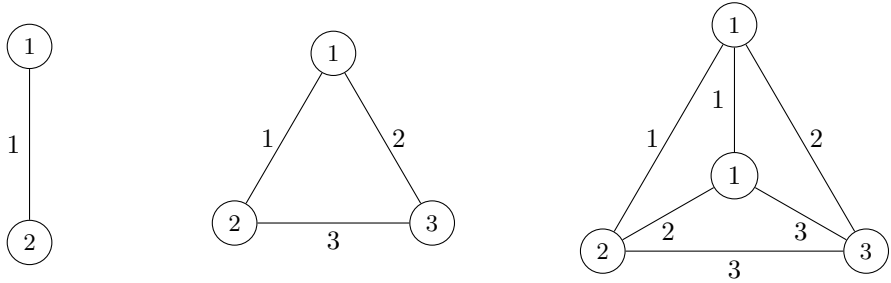
\begin{figure} %[htb]
\begin{center}
\begin{minipage}[c]{.2\linewidth}
\begin{tikzpicture}
\def \n {2}
\def \radius {1.3cm}
\def \nradius {0.6cm}
\def \margin {8} % margin in angles, depends on the radius

\foreach \s in {1,...,\n}
{
\pgfmathparse{\s} \let \z \pgfmathresult;
  \node (\s) at ({360/\n * (\s - 1) + 90}:\radius) [draw,circle, fill=none, minimum size=0.5cm] {\small \pgfmathprintnumber{\z}};
}

\draw (1)  -- (2)  node [midway, left, fill=none] {$1$};
\end{tikzpicture}
\end{minipage}
\begin{minipage}[c]{.35\linewidth}
\begin{tikzpicture}

\def \n {3}
\def \radius {1.5cm}
\def \nradius {0.6cm}
\def \margin {8} % margin in angles, depends on the radius

\foreach \s in {1,...,\n}
{
\pgfmathparse{\s} \let \z \pgfmathresult;
  \node (\s) at ({360/\n * (\s - 1) + 90}:\radius) [draw,circle, fill=none, minimum size=0.5cm] {\small \pgfmathprintnumber{\z}};

}

\draw (1)  -- (2)  node [midway, left, fill=none] {$1$};
\draw (1)  -- (3)  node [midway, right, fill=none] {$2$};
\draw (2)  -- (3)  node [midway, below, fill=none] {$3$};
\end{tikzpicture}
\end{minipage}
\begin{minipage}[c]{.35\linewidth}
\begin{tikzpicture}

\def \n {3}
\def \radius {2cm}
\def \nradius {0.6cm}
\def \margin {8} % margin in angles, depends on the radius

\foreach \s in {1,...,\n}
{
\pgfmathparse{\s} \let \z \pgfmathresult;
  \node (\s) at ({360/\n * (\s - 1) + 90}:\radius) [draw,circle, fill=none, minimum size=0.5cm] {\small \pgfmathprintnumber{\z}};

}

 \node (0) at (0,0) [draw,circle, fill=none, minimum size=0.5cm] {$1$};

\draw (1)  -- (2)  node [midway, left, fill=none] {$1$};
\draw (1)  -- (3)  node [midway, right, fill=none] {$2$};
\draw (0)  -- (1)  node [midway, left, fill=none] {$1$};
\draw (0)  -- (2)  node [midway, below, fill=none] {$2$};
\draw (2)  -- (3)  node [midway, below, fill=none] {$3$};
\draw (0)  -- (3)  node [midway, below, fill=none] {$3$};
\end{tikzpicture}
\end{minipage}

\caption{\label{smallKn} Equitable  total nsd-colourings of $K_2$, $K_3$ and $K_4$.}
\end{center}
\end{figure}

%\begin{pf} (of Thm.~\ref{TheoremEquitableSumTotalComplete})
 The colourings of small cases (for $n\leq 4$) are depicted in Figure~\ref{smallKn}.
By Theorem~\ref{TheoremEquitableSumCompleteGraph} and Proposition~\ref{prop:total_edges},
it is enough to show that  $\EchiT(K_n)\neq 2$ for $n\geq 3$
(recall that $\EchiT(K_n)\ge 2$ since $K_n$ is regular).

 First we observe that for any $n$, there exist (essentially) exactly two total $2$-colourings distinguishing $K_n$ by sums; in one of them there is a vertex with a monochromatic palette of $1$'s, and in the other one there is a vertex with a monochromatic palette of $2$'s, since we add $n$ numbers (colours) at every vertex and we obtain sums from the set $\{n, ..., 2n\}$ but not $n$ and $2n$ in the same colouring.
 Moreover, it is easy to observe that for any such colouring of $K_n$, $n\geq 2$, with a monochromatic palette of $a$'s, $a\in\{1,2\}$, say at a vertex $v$, after deleting $v$ we obtain a  total $2$-nsd-colouring of $K_{n-1}$ with a monochromatic palette of $(3-a)$'s. 
 %if there exists a total distinguishable colouring for $K_n$ with a vertex $v$ having a monochromatic palette of %$1$'s, then in $K_{n-1}$ there exists a vertex $u$ with a monochromatic palette of $2$'s, since the colouring is %distinguishing and $u$ and $v$ have different sets of colours in $K_n$.

Let $\gamma$ be such a total $2$-colouring distinguishing  $K_n$, $n\geq 2$, by sums with a vertex having a monochromatic palette of $1$'s - for a monochromatic palette of $2$'s the reasoning and calculations are the same. Now for every positive integer $k$ we prove by induction that:
{\it If $n=2k$ then there exist $k(k+1)$ elements (vertices and edges) coloured with $1$ and $k^2$ elements coloured with $2$.
If $n=2k+1$ then there exist $(k+1)^2$ elements coloured with $1$ and $k(k+1)$ elements coloured with $2$. }

For $n=2$ we have both vertices in different colours and an edge is coloured with $1$, since there exists a monochromatic palette of $1$'s. For $n=2k$, $k>1$, let $v$ be a vertex with a monochromatic palette of $1$'s. Then there exists a vertex $u$ in $K_{n-1}=K_n-v$ having a monochromatic palette of $2$'s. Consider a graph $K_{2k-2}$ obtained from $K_{2k}$ by removing vertices $u$ and $v$. Then there exist $k(k-1)$ elements coloured with $1$ and $(k-1)^2$ elements coloured with $2$ in $K_{2k-2}$, by induction. So, there exist $k(k-1)+ 2k$ elements coloured with $1$ (since $v$ adds $n$ elements coloured with $1$) and $(k-1)^2+2k-1$ elements coloured with $2$ (since $u$ adds $n-1$ elements coloured with $2$)  in $K_{2k}$.

Similarly, for $n=3$ we have four elements coloured with $1$ and two elements coloured with $2$, since there exists a monochromatic palette of $1$'s and there exist exactly one such a total distinguishable colouring by sums. For $n=2k+1$, $k>1$, let $v$ be a vertex with a monochromatic palette of $1$'s. Then there exists a vertex $u$ in $K_{n-1}=K_n-v$ having a monochromatic palette of $2$'s. Consider a graph $K_{2k-1}$ obtained from $K_{2k+1}$ by removing vertices $u$ and $v$. Then there exist $k^2$ elements coloured with $1$ and $k(k-1)$ elements coloured with $2$ in $K_{2k-1}$, by induction. So, there exist $k^2+ 2k+1$ elements coloured with $1$ (since $v$ adds $n$ elements coloured with $1$) and $k(k-1)+2k$ elements coloured with $2$ (since $u$ adds $n-1$ elements coloured with $2$)  in $K_{2k+1}$.

Now observe that if $n=2k$ then a difference between numbers of $1$'s and $2$'s is $k$ in the colouring $\gamma$ and if $n=2k+1$ the difference is $k+1$. So, $\gamma$ is equitable only for $n=2$. This finishes the proof, as the same reasoning applies in the case of a monochromatic palette of $2$'s in $\gamma$ (with $1$'s and $2$'s switched).

%\qed
%\end{pf}

%%%%%%%

\section{Discussion}

We introduced and studied in this paper equitable  edge and total nsd-colourings.
We determined the equitable  nsd-index
of complete graphs (Theorem~\ref{TheoremEquitableSumCompleteGraph}), 
complete bipartite graphs (Theorem~\ref{TheoremEquitableSumCompleteBipartiteGraph})
 and forests (Theorem~\ref{TheoremEquitableSumForest}),
and the equitable  total nsd-chromatic number
of bipartite graphs (Theorem~\ref{TheoremEquitableSumTotalBipartite}) 
and complete graphs (Theorem~\ref{TheoremEquitableSumTotalComplete}).

By colouring the edges of a graph $G$ (having no isolated edge) with different
powers of~2, one obviously get an equitable  edge nsd-colouring of $G$,
so that the inequality $\EchiE(G)\le 2^{|E(G)|}$ holds for every graph $G$
with no isolated edge.
In a recent paper~\cite{BSL16}, Bensmail, Senhaji and Szabo Lyngsie studied
``edge-injective''  nsd-colourings,
a stronger version of equitable  edge nsd-colourings
in which no two edges can be assigned the same colour.
From their results, it follows that 
the inequality 
$$\EchiE(G)\le \min\left(2|E(G)|,|E(G)|+2\Delta(G)\right)$$ 
holds for every graph $G$
(having no isolated edge) with maximum degree $\Delta(G)$, giving an upper bound on $\EchiE(G)$
which is polynomial in terms of $|E(G)|+|V(G)|$.
However, we do not know if there exists any constant upper bound on
$\EchiE(G)$ for every graph $G$.

Theorem~\ref{TheoremEquitableSumTotalBipartite} shows that 
$\EchiT(G)\le 2$ holds for every bipartite graph $G$.
It would be interesting to determine whether there also
exists a constant upper bound on $\EchiE(G)$ when $G$ is bipartite.

\section*{Acknowledgements}

The research of all the authors except the third one were supported by the joint research program PICS CNRS 6367 ``Graph partitions''.
Most of this work was done while O.~Baudon, M. Senhaji and \'E.~Sopena
were visiting AGH University of Science and Technology, and while
M. Pil\'sniak, J.~Przyby{\l}o and M.~Wo\'zniak were visiting LaBRI.

J.~Przyby{\l}o was supported by the National Science Centre, Poland, grant no. 2014/13/ B/ST1/01855 while 
M.~Wo\'zniak was supported by the National Science Centre, Poland, grant no. DEC-2013/09/B/ST1/01772.
Moreover, M. Pil\'sniak, J.~Przyby{\l}o and M.~Wo\'zniak were
partly supported by the Polish Ministry of Science and Higher Education,
while \'E.~Sopena was partly supported by the Cluster of excellence CPU, from the
Investments for the future Programme IdEx Bordeaux (ANR-10-IDEX-03-02).


\section*{References}

\begin{thebibliography}{99}

\bibitem{Louigi30}
L. Addario-Berry, K. Dalal, C. McDiarmid, B.A. Reed, A. Thomason.
Vertex-Colouring Edge-Weightings.
 {\em Combinatorica} 27:1 (2007), 1--12.

\bibitem{Louigi}
L. Addario-Berry, K. Dalal, B.A. Reed. 
Degree Constrained Subgraphs.
{\em Discrete Appl. Math.} 156:7 (2008), 1168--1174.

\bibitem{BSL16}
J.~Bensmail, M.~Senhaji, K.~Szabo Lyngsie.
Equitability, edge-injectivity, and the 1-2-3 Conjecture.
Preprint (2016), available at \url{https://hal.archives-ouvertes.fr/hal-01361482}.

\bibitem{CLWY11} G.J. Chang, C. Lu, J. Wu, Q. Yu.
Vertex-coloring edge-weightings of graphs.
{\em Taiwanese J. Math.} 15:4 (2011), 1807--1813.

\bibitem{Erdos}
P. Erd\"os. 
Problem 9.
In {\it Theory of Graphs and Its Applications} (M. Fieldler, Ed.),
159, Czech. Acad. Sci. Publ., Prague, 1964.

\bibitem{HS70}
A. Hajnal and E. Szemer\'edi.
Proof of a conjecture of P. Erd\"os.
In {\it Combinatorial
Theory and its Application} (P. Erd\"os, A. R'enyi, and V. T. S'os, Eds.), pp. 601-623,
North-Holland, London, 1970.

\bibitem{Kalkowski12}
M. Kalkowski.
A note on 1,2-Conjecture.
%{\em Electron. J. Combin.}, to appear.
In Ph.D. Thesis, 2009.

\bibitem{KalKarPf123}
M. Kalkowski, M. Karo\'nski, F. Pfender.
Vertex-coloring edge-weightings: Towards the 1-2-3 conjecture.
{\em J. Combin. Theory, Ser. B} 100 (2010), 347--349.

\bibitem{123KLT}
M. Karo\'nski, T. \L uczak, A. Thomason.
Edge weights and vertex colours.
{\em J. Combin. Theory Ser. B} 91 (2004), 151--157.

\bibitem{KNNSS12} M. Khatirinejad, R. Naserasr, M. Newman, B. Seamone, B. Stevens.
Vertex-colouring edge-weightings with two edge weights.
{\em Discrete Math. Theoret. Comput. Sci.} 14:1 (2012), 1--20.

%\bibitem{KK12}
%H.A.~Kierstead and A.V.~Kostochka.
% Every 4-colorable graph with maximum degree 4 has an equitable 4-coloring.
% {\em J. Graph Theory} 71 (2012), 31--48.

\bibitem{KK08}
H.A. Kierstead and A.V. Kostochka. 
A short proof of the Hajnal-Szemer\'edi Theorem on equitable coloring.
{\it Combinatorics, Probability and Computing} 17 (2008), 265--270.


\bibitem{L13}
K.-W. Lih.
 Equitable coloring of graphs.
 In {\em  Handbook of Combinatorial Optimization}, 2nd ed., 
 P.M.~Pardalos, D.-Z.~Du and R.L.~Graham (Editors), Springer, New
York (2013), 1199--1248.

\bibitem{LYZ11} H. Lu, Q. Yu, C.-Q. Zhang.
Vertex-coloring 2-edge-weighting of graphs.
{\em Europ. J. Combin.} 32 (2011), 21--27.

%\bibitem{Prz08} J. Przyby{\l}o.
%A Note on Neighbour-Distinguishing Regular Graphs Total-Weighting.
%{\em Electr. J. Comb.} 15 (2008), \#N35.

\bibitem{12Conjecture}
J. Przyby{\l}o, M. Wo\'zniak.
On a 1,2 Conjecture.
{\em Discrete Math. Theoret. Comput. Sci.} 12:1 (2010), 101--108.

%\bibitem{Survey2012} B. Seamone.
%The 1-2-3 Conjecture and related problems: a survey.
%\href{http://arxiv.org/abs/1211.5122}{CoRR abs/1211.5122} (2012).

\bibitem{123with13}
T. Wang, Q. Yu. 
On vertex-coloring 13-edge-weighting.
{\em Front. Math. China}
3:4 (2008), 581--587.

\end{thebibliography}
\end{document}